\documentclass[11pt,a4paper]{article}
\usepackage{jheppub}
\usepackage{amssymb,amsmath,amsfonts,mathrsfs}
\usepackage{graphicx}
\usepackage[dvipsnames]{xcolor}
\usepackage{verbatim}
\usepackage{epsfig}
\usepackage{color}
\usepackage{multirow}
\usepackage{comment}
\usepackage{datetime}
\usepackage{slashed}
\usepackage{ulem}
\usepackage{framed}
\usepackage{longtable}
\usepackage[mathscr]{euscript}
\usepackage{cancel}
\usepackage{tensor}
\usepackage{mathtools}
\usepackage{caption}
\usepackage{subcaption}
\usepackage[multiple]{footmisc}
\usepackage{pgfplots}
\usepackage{natbib}
\usepackage[applemac]{inputenc}
\usepackage{tikz}
\usepackage{placeins}
\usetikzlibrary{decorations.markings}

  \usepackage{multirow}

\newcommand{\ft}[2]{{\textstyle\frac{#1}{#2}}}
\newcommand{\nn}{\nonumber}

\def\be{\begin{equation}}
\def\ee{\end{equation}}
\def\bea{\begin{align}}
\def\eea{\end{align}}
\def\beaq{\begin{eqnarray}}
\def\eeaq{\end{eqnarray}}

\def\a{\alpha} \def\b{\beta} \def\g{\gamma} 
   \def\q{\theta}
 \def\k{\kappa}

\title{On the stability and deformability of top stars}
\author{Massimo~Bianchi,}
\author{Giorgio~Di~Russo,}
\author{Alfredo~Grillo,}
\author{Jose~Francisco~Morales,}
\author{Giuseppe~Sudano}
\affiliation{Dipartimento di Fisica, Universit\`a di Roma ``Tor Vergata" \& Sezione INFN Roma2, Via della ricerca scientifica 1, 00133, Roma, Italy}

\abstract{Topological stars, or top stars for brevity, are smooth horizonless static solutions of Einstein-Maxwell theory in 5-d that  reduce to  spherically symmetric solutions of Einstein-Maxwell-Dilaton theory in 4-d. We study linear scalar perturbations of top stars and argue for their stability and deformability. We tackle the problem with different techniques including WKB approximation, numerical analysis, Breit-Wigner resonance method and quantum Seiberg-Witten curves. We identify three classes of quasi-normal modes corresponding to prompt-ring down modes, long-lived meta-stable modes and what we dub `blind' modes. All mode frequencies we find have negative imaginary parts, thus suggesting linear stability of top stars. Moreover we determine the tidal Love and dissipation numbers encoding the response to tidal deformations and, similarly to black holes, we find zero value in the static limit but, contrary to black holes, we find non-trivial dynamical Love numbers and vanishing dissipative effects at linear order. For the sake of illustration in a simpler context, we also consider a toy model with a piece-wise constant potential and a centrifugal barrier that captures most of the above features in a qualitative fashion. }

\begin{document}
\maketitle
\flushbottom 
\section{Introduction}
Finding smooth horizonless solutions with the same mass, charges and angular momenta of `putative' black holes is one the crucial step in the fuzzball proposal, according to which (semi)classical horizon-full BHs emerge from coarse-graining of micro-state geometries \cite{Lunin:2001fv, Mathur:2009hf}. 
In specific cases (a)typical micro-states admit a geometric description at low-energy as supersymmetric smooth horizonless solutions to Einstein equations coupled to scalars, vectors and other p-forms (in higher dimensions) \cite{Bianchi:2017bxl, Bena:2017upb, Turton:2012ny, Giusto:2011fy, Bianchi:2016bgx}.

Given the enormous success in the context of supersymmetric 2-charge black strings, 3- and 4-charged BHs in four and five dimensions \cite{Bena:2015bea, Bena:2016agb, Bena:2016ypk, Bena:2017xbt}, recent attempts were made to extend the analysis to non-supersymmetric and neutral cases \cite{Bah:2020pdz}. While JMaRT solutions \cite{Jejjala:2005yu} have been shown to suffer an ergo-region instability \cite{Cardoso:2005gj} as well as a `charge instability' \cite{Bianchi:2023rlt}, the stability of special solutions recently constructed by Bah and Heidmann dubbed `topological stars' \cite{Bah:2020pdz, Heidmann:2021cms, Bah:2022pdn, Heidmann:2022zyd} (or `top stars' for brevity in the present investigation) has been established at the classical and thermodynamical level in \cite{Miyamoto:2007mh, Stotyn:2011tv, Bah:2021irr}. Moreover their imaging based on geodesics has been studied in \cite{Heidmann:2022ehn}.

Topological stars, or top stars for brevity, are smooth horizonless static solutions of Einstein-Maxwell theory in 5-d that admit a reduction to black hole solutions of Einstein-Maxwell-Dilaton theory in 4-d.
They have been devised in the attempt to intertwine between the `bottom-up' and the `top-down' approaches in the construction of micro-state geometries. Within the former, smooth horizonless exotic compact objects (ECOs) have been constructed, whose observables differ from those of black holes in GR \cite{Bianchi:2020bxa, Bianchi:2020miz, Bena:2020see, Bena:2020uup, Bah:2021jno, Thorne:1980ru, Hansen:1974zz, Bena:2007kg}. Still, generic 4-dimensional ECOs tend to lack a consistent UV completion and typically require the introduction of exotic matter or mechanisms in order to bypass the no-hair theorem, which would otherwise forbid the replacement of the horizon with other structures. 
The framework of ten-dimensional Superstring Theory, instead,  provides a  top-down approach that bypasses no-hair theorem and information paradox problems by embedding BHs into higher dimensional smooth horizonless geometries with structure at horizon scales built out of ordinary matter. The fuzzball proposal has been formulated within this approach in  \cite{Lunin:2001jy, Mathur:2009hf}. However, the solutions are typically supersymmetric, and when they are not, they tend to be (over-)rotating \cite{Jejjala:2005yu} and involve a the large amount of fields making the extraction of relevant phenomenological observables cumbersome \cite{Cardoso:2005gj, Bianchi:2019lmi, Bianchi:2023rlt}. On the other hand, top stars are smooth horizonless solutions of Einstein-Maxwell theory, where Gregory-Laflamme instability \cite{Gregory:1993vy} is prevented by the presence of electro-magnetic fluxes. 

Aim of the present paper is to study the response of top stars to scalar perturbations. Due to spherical symmetry in 4-d, non-trivial dynamics arises only along the radial direction and it is governed  by a Confluent Heun Equation (CHE) similar to the one governing all non-extremal BHs in 4 dimensions \cite{Bianchi:2021mft, Bonelli:2021uvf, Bonelli:2022ten, Consoli:2022eey, Fioravanti:2020udo, Fioravanti:2021bzq, Fioravanti:2021dce, Dodelson:2022yvn}. We probe solutions by classical scattering (geodesics) and waves. The spectrum of Quasi Normal Modes (QNMs) gives precious information not only on the stability of the solution but also on the (gravitational) wave signal produced, for instance, in the ring-down phase of a binary coalescence that leads to a stable remnant \cite{Ikeda:2021uvc}.

We will exploit the newly proposed correspondence between perturbation theory in gravity and quantum Seiberg-Witten (qSW) curves \cite{Bianchi:2021mft}. In this framework, all non-trivial information about the radial dynamics is coded in a single holomorphic function, the prepotential ${\cal F}(a)$,
describing the dynamics of a  ${\cal N}=2$ supersymmetric $SU(2)$ gauge theory with three fundamental hypermultiplets.  

Our analysis suggests that top stars are stable at the linearized level at least for (massless and massive) scalar perturbations, that can carry charge in the form of KK momentum. 
We cannot exclude that different perturbations (e.g. winding states or wrapped branes) don't lead to instability. Even more so, we cannot exclude non-linear instabilities \cite{Eperon:2016cdd} that plague some microstate geometries, including BPS ones \cite{Giusto:2004id, Giusto:2004ip} with `evanescent' ergo-sphere.

Using qSW techniques, we will also determine the response of top stars to tidal deformations and show that Tidal Love Numbers (TLNs) vanish in the static limit $\omega=0$ but are non-zero in  dynamical regimes with  $\omega\neq 0$. However, contrary to BHs, whose static TLNs also vanish in $D=4$ \cite{Hui:2020xxx, Chia:2020yla}, the dynamical TLNs turn out to be real rather than purely imaginary, signalling a `real' deformability rather than a dissipative response.

The plan of the paper is as follows.
In Section 2 we describe a toy model with a piecewise constant potential and a centrifugal barrier that captures qualitative features of BH and fuzzball perturbations.   
In Section 3 we briefly review the geometry of top stars.  Then we study geodesics, exploiting the trivial separation of the dynamics. In particular we focus on critical geodesics that form the so-called light-rings and estimate the spectrum of QNMs based on a WKB approximation of the dynamics. The results will be later used as a seed for the numerical and qSW analysis of the wave equation. 
In section 4 we study QNMs and tidal response using numerical methods.  In Section 5 we compute QNMs and tidal responses of top stars using qSW techniques. We show that, at the linearized level in the `probe' approximation, neither absorption nor (charge) super-radiance can take place, in a smooth geometry such as a top star's. 
Section 6 draws our conclusions and contains a discussion of the stability at linearized level in view of possible `phenomenological' implications for this kind of smooth horizonless geometries or the more appealing but by-far more challenging `Schwarzschild topological solitons' \cite{Bah:2022yji, Bah:2023ows} or `Bubble Bag End geometries' \cite{Bah:2021owp, Bah:2021rki}.

\section*{Note added}
While we were completing this project, we were informed by Pierre Heidmann that he was performing a similar analysis in collaboration with Ibrahima Ba, Emanuele Berti, Nicholas Speeney \cite{cavity}.

\section{A toy model for wave dynamics}

Linearized perturbations around classical BHs and some, particularly symmetric, fuzzball geometries are often captured by Schr\"odinger-like equations governing the dynamics along radial and angular directions. For spherically symmetric BHs one gets the Regge-Wheeler-Zerilli equations \cite{Regge:1957td, Zerilli:1970se}, while for rotating BHs one gets the Teukolsky equations \cite{Teukolsky:1972my}. Thanks to the large (hidden) isometry, the dynamics separates and in both cases one ends up with second order ordinary differential equation, that can  be put in canonical Schr\"odinger-like form 
  \be
  \Psi''(x) +Q(x) \Psi(x)=0 \label{sch1}
  \ee
describing the quantum mechanics\footnote{Despite similarity in the wording and nomenclature and even when we will introduce `$\hbar$' to `quantize' Seiberg-Witten curves, everything we say pertains to the classical regime in gravity.} of a particle of energy $E$ in a potential $V(x)$ with 
$Q(x)=E-V(x)$. In the context of gravity, $E= \omega^2$ with $\omega$ the frequency of the wave and $Q(x)=P_x^2$ parametrises the radial or angular momenta carried by the wave.  For non-rotating solutions $Q(x)$ is given by a polynomial of degree-two in $\omega$, while in the case of rotating BHs and fuzzballs, a further dependence arises from (Carter's) separation constant introduced to disentangle radial and angular motion.

A zero of $Q(x)$ represents a turning point of geodetic motion. For specific choices of the frequency $\omega_c$, such that 
the zero $x_c$ of $Q$ is also a zero of its first derivative, the time to reach the turning point $x_c$ (or to scape from it)  is infinite. The resulting critical geodesics behaves as an attractor, i.e. a {\it light ring},  for probes with the prescribed frequency. A light ring is then defined by the conditions
\be
 Q(\omega_c,x_c)=Q'( \omega_c, x_c)=0 \label{lightringeq}
\ee
 There are two types of light rings depending on the sign of the second derivative of $Q$. 
For $ Q''(x_c)<0$, the light ring is said to be (meta)stable, while a light ring is said to be unstable otherwise. 
  
  Aim of this paper is to probe top stars with particles and waves, exploiting the separable dynamics whereby radial motion is described by
 a Schr\"odinger-like equation of type (\ref{sch1}). We will deal with three quantities characterizing the response of the geometry: the QNMs, the tidal Love numbers (TLNs) and the Dissipative numbers (DNs). 
 
QNMs are solutions of (\ref{sch1}) satisfying certain boundary conditions in the interior (ingoing waves for BHs at the horizon, and regularity
 for smooth geometries at the cap) and at infinity (typically outgoing waves). If we denote by $\Psi_{\rm in}(x)$ and $\Psi_{\rm out}(x)$ some solutions with the prescribed boundary conditions in the interior and at infinity respectively, QNMs are defined by the condition that the Wronskian between the two solutions vanishes: 
  \be
  W_{ {\rm in} , {\rm out} }=\Psi_{\rm in}(x)\, \Psi_{\rm out}'(x) - \Psi'_{\rm in}(x)\, \Psi_{\rm out}(x) =0 \label{wrong}
  \ee 
We notice that, in general, for ordinary differential equations of  second order, the vanishing of the Wronskian at a point implies the vanishing everywhere, so this condition is $x$-independent\footnote{In particular for equations in the form (\ref{sch1}) the Wronskian is constant.} and becomes and eigenvalue equation for $\omega$. 
  Equation (\ref{wrong}) follows from imposing the matching of the two functions and their first derivatives at a point $x$, and it has an infinite number of solutions in the complex $\omega$-plane, the {\it QNM frequencies} $\omega_n$. 
   
   On the other hand, the TLNs $L$ and the dissipative number $\Theta$, describe the response of the geometry to tidal perturbations. They are defined as the real and imaginary parts of the ratio between the coefficients of the term $r^{-\ell-1}$ (response) and the one $r^\ell$ (source), of the wave function in the intermediate region (near zone), far away from the horizon/cap, but still close enough to  feel centrifugal forces dominating over the gravitational potential.  
   We write
     \be\label{wavfunc10}
\Psi(x) =\left\{\begin{array}{lll}
        A \, \Psi_{\rm in}  (x) &~~~~~~~~&   {\rm Cap/Horizon}\\
        B\, \Psi_{\rm resp}  (x) +   C \,  \Psi_{\rm source}  (x)   &&   {\rm Near ~ Zone}\\
            \end{array} \right.
 \ee
 with $ \Psi_{\rm resp}  (x) $ and  $\Psi_{\rm source}  (x)$ the dying and growing solutions respectively in the near zone. 
   Imposing the left-right matching of  $\partial_x \ln \Psi(x)$ one finds the coefficients ratio  
     \be
  {\cal L}= L+{\rm i} \Theta={B \over C} =- { W_{ {\rm in} , {\rm source} } \over W_{ {\rm in} , {\rm resp} } }
  \ee
   We notice that $\Psi_{\rm resp} $ and $\Psi_{\rm source}$ are real functions, so dissipative effects appears only when $\Psi_{\rm in}$ is complex, i.e. when ingoing or outgoing fluxes are allowed at the boundary (horizon).  
 
In many known cases the relevant Schr\"odinger-like equations can be put in the form of a Heun Equation (or its confluences), where $Q(x)$ has generically four singular points with poles of second order. For (A)dS Kerr-Newman BHs in four dimensions the four  singularities represent the four (complex) horizons.  
On the other hand, linearized perturbations around all non-extremal BHs in $D=4$ with flat asymptotics are governed by some Confluent Heun Equation (CHE), that becomes a Doubly Confluent Heun Equation (DCHE) in the extremal case \cite{Aminov:2020yma, Bianchi:2021mft, Bianchi:2021xpr, Bonelli:2021uvf, Bonelli:2022ten, Consoli:2022eey}.  The relevant solution to these HEs can be elegantly related to observables in four-dimensional ${\cal N}=2$ Super Yang-Mills theory (SYM) living on Nekrasov-Shatasvili (NS) $\Omega$ background \cite{Nekrasov:2009rc} or, thanks to AGT (Alday Gaiotto Tachikawa) correspondence \cite{Alday:2009aq}, to degenerate five-point correlators in two-dimensional Liouville theory. The whole non-triviality of the dynamics is coded in a single holomorphic function: the gauge theory prepotential \cite{Seiberg:1994aj, Seiberg:1994rs, Bruzzo:2002xf, Flume:2002az, Nekrasov:2003af}. Still, this function cannot be computed, in general, in a closed analytical form, so it is often convenient to  consider some simplified model that capture qualitative features of the dynamics. This is what we are going to do next.

\subsection{Piece-wise constant potentials with centrifugal barrier}

In this section, we discuss a simple toy model for wave dynamics that captures many of the essential features of QNM dynamics  and tidal responses. We consider a potential of the form
\be
Q(x)=\omega^2-V_{\rm eff}(x) =\omega^2 -V_{\rm pwc}(x) -{\ell(\ell+1)\over  x^2 }
\ee
with the first piece-wise constant term $V_{\rm pwv}(x)$ representing a `caricature' of the interaction potential (gravitational or electromagnetic) and the last $\ell$-dependent term  representing the centrifugal barrier. 
We take  $V_{\rm pwc}(x)$ of the form
\be
V_{\rm pwc} (x) = \left\{\begin{array}{lcl}
         V_1  &~~~~ &  0\le x\leq L_1\\
         V_2 & & L_1< x\leq L_2\\
          0  & & x>L_2
        \end{array} \right. 
 \ee
\begin{figure}[htbp]
\begin{minipage}{0.4\textwidth}
\centering
\includegraphics[width=1\textwidth]{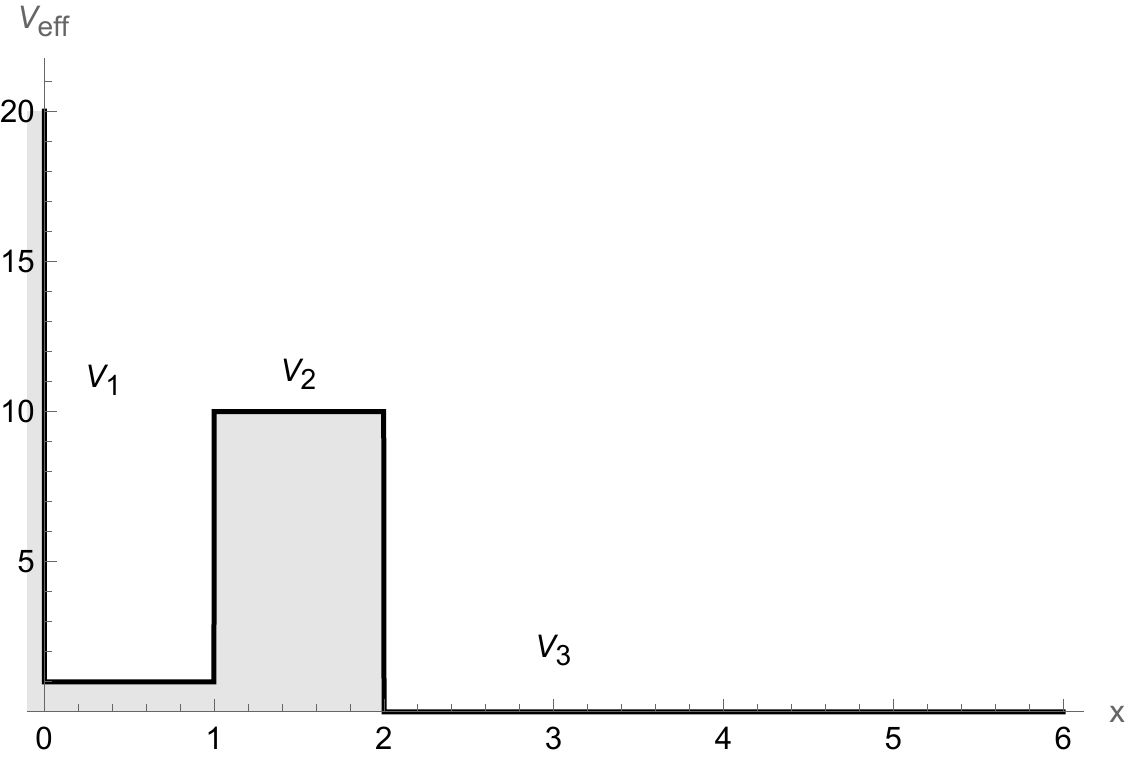}
\end{minipage}
\hspace{2cm}
\begin{minipage}{0.4\textwidth}
\centering
\includegraphics[width=1\textwidth]{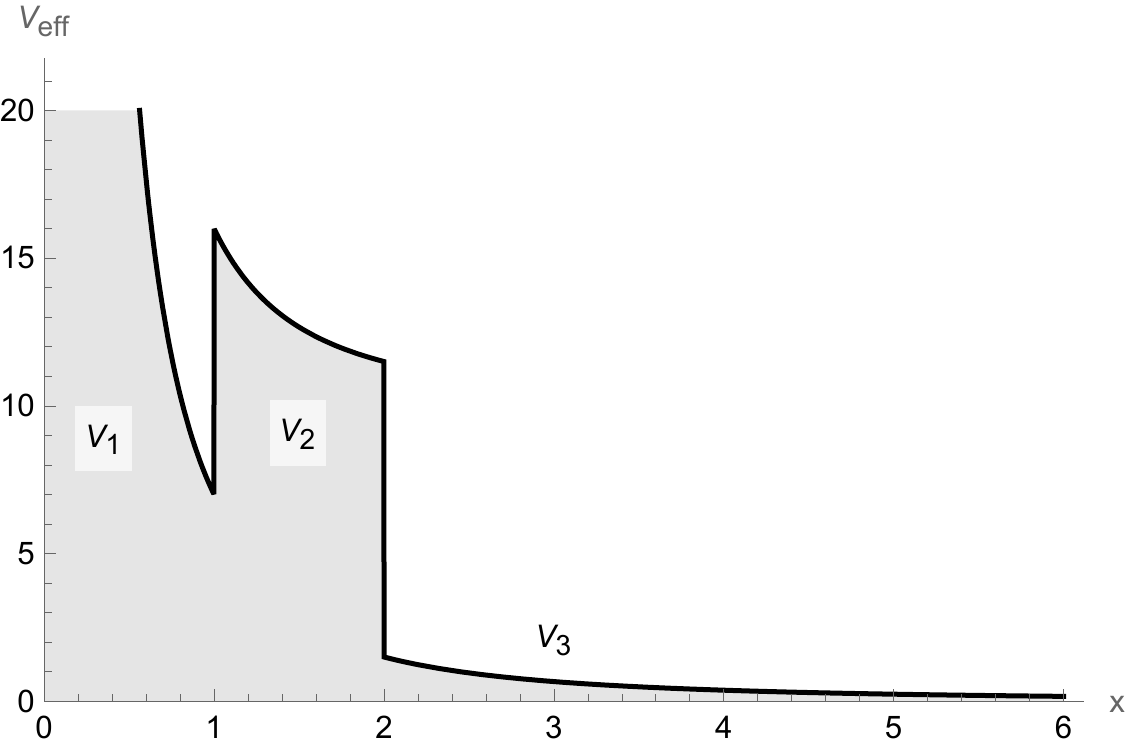}
\end{minipage}
\caption{ Effective potential $V_{\rm eff}(x)$ }\label{graphVeff}
\end{figure}

The effective potential is displayed in figure \ref{graphVeff}.  The important difference with respect to the effective potential of BH solutions is the existence of a minimum and a maximum, i.e. a (meta)stable and an unstable light ring\footnote{Without much effort one could include a stable minimum, representing a stable orbit, that would mimic an innermost circular stable orbit (ISCO), where the accretion disk is located.}.

 We will mainly focus on the interval $V_1 < \omega^2 < V_2$ and introduce the quantum numbers
 \be
   k_1=\sqrt{\omega^2-V_1} \quad , \quad \beta=\sqrt{V_2-\omega^2} \quad , \quad k_3=\omega
 \ee


\subsection{QNMs for $\ell=0$}

 We start by considering the s-wave $\ell=0$ that corresponds to keeping only the piece-wise constant part of the potential.  We can consider various boundary conditions at the interior: Neumann, Dirichlet, ingoing waves.
 Here we compute quasi-normal modes and tidal Love and dissipation numbers for the various choices.  

\subsection*{The infinite wall}

\begin{figure}[t]
\begin{minipage}{0.25\textwidth}
\centering
\includegraphics[width=1\textwidth]{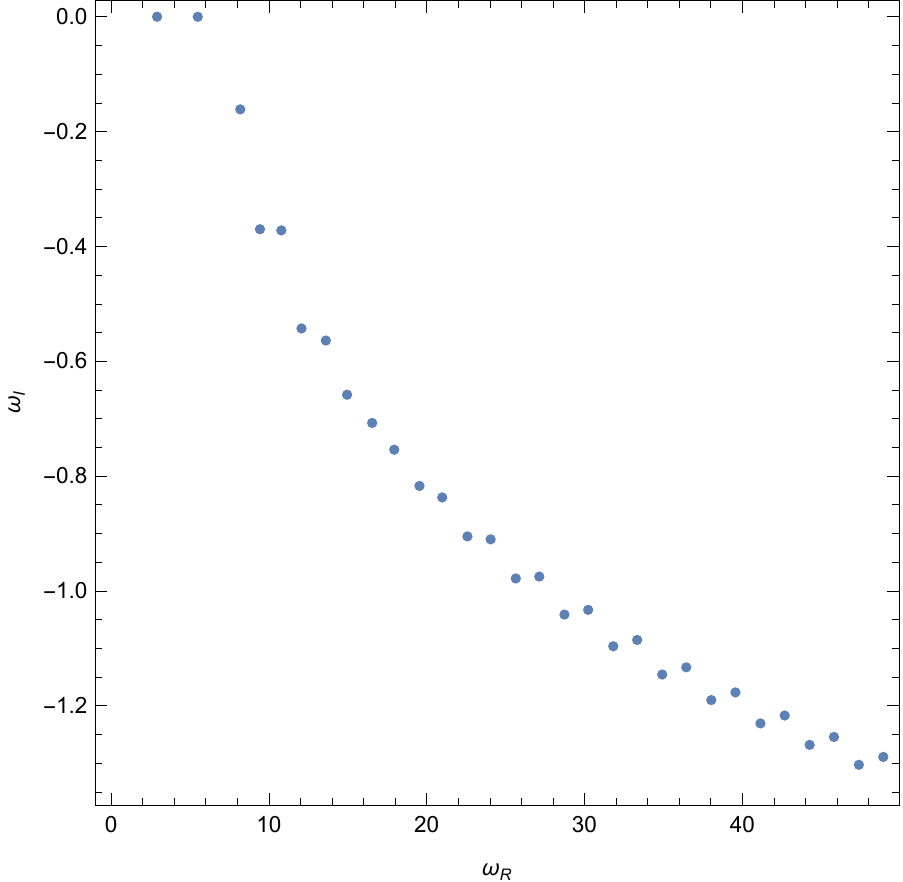}
\end{minipage}
\hspace{0.2cm}
\begin{minipage}{0.25\textwidth}
\centering
\includegraphics[width=1\textwidth]{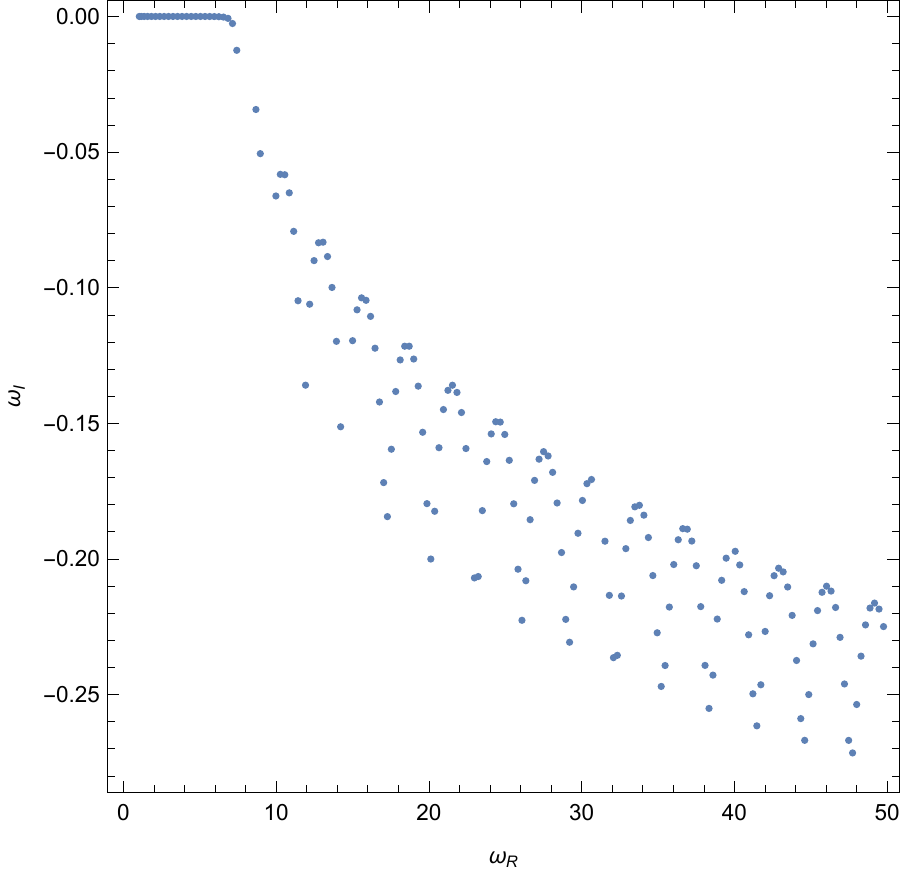}
\end{minipage}
\hspace{0.2cm}
\begin{minipage}{0.4\textwidth}
\centering
\includegraphics[width=1\textwidth]{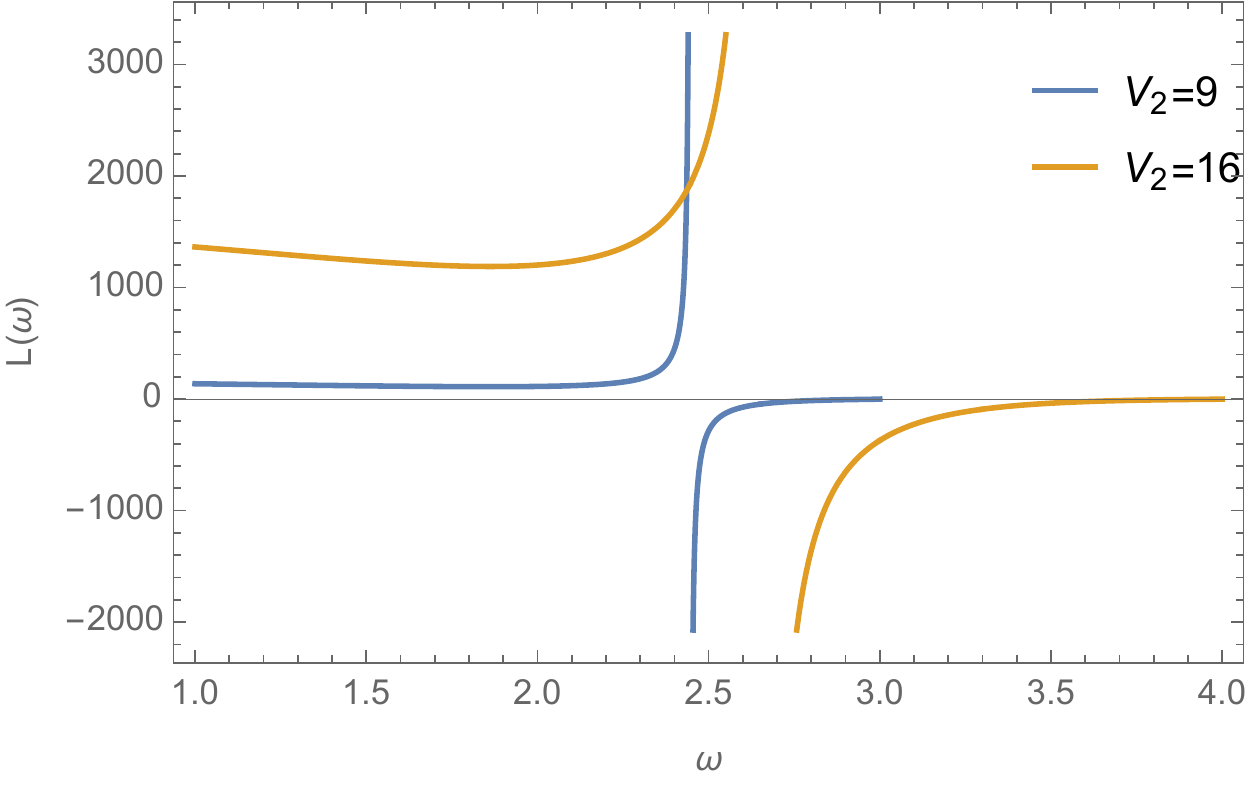}
\end{minipage}
\caption{QNMs and Tidal Love numbers for $\ell=0$ waves with Dirichlet boundary conditions: a) QNMs for square potential barriers with
$V_1=1$, $V_2=50$, $\Delta=1$, $L_1=1$. b)  $L_1=10$. c) Love numbers $V_1=\Delta=L_1=1$.}\label{qnmplanefull}
\end{figure}

   In presence of an infinite wall, the wave function vanishes at $x=0$ and we have 
 \be\label{wavfunc2}
\Psi^{\rm Wall}(x) =\left\{\begin{array}{lll}
        A_0 \sin k_1 x, &~~~~~~~~&   0\leq x\leq L_1\\
        B_0 e^{-\beta x}+   C_0 e^{\beta x}  &&   L_1< x\leq L_2\\
         D_0 e^{i k_3 x} +\tilde{D}_0 e^{-i k_3 x}  &&   x>L_2
        \end{array} \right.
 \ee
  Although $\ell=0$, one can still formally define the Love number as the ratio of the $B_0$ and $C_0$ coefficients. 
Matching the functions in (\ref{wavfunc2})  and its first derivative on the left and on the right of $x=L_1$ and $x=L_2$  one finds
\be
{\cal L}^{W}_0(\omega)={B_0(\omega)\over C_0(\omega)} = {\b \tan k_1 L_1 - k_1 \over \b \tan k_1 L_1 + k_1} e^{2 \b L_1}
\ee
The results for some choices of the potentials are shown in figure \ref{qnmplanefull}.  The function ${\cal L}$ is real as expected (no dissipation) and exhibits infinite number of poles on the real $\omega$-line given by the solutions of the transcendental eigenvalue equation
\be
 \tan k_1 L_1 =-{k_1\over \beta} \label{metalove}
\ee
 They correspond to metastable modes localized inside the  cavity. 
QNMs correspond to solutions with $\tilde{D}_0=0$. Imposing the matching of the function and its derivatives at $L_1$ and $L_2$ leads to the eigenvalue equation
\be
\frac{\beta}{k_1}\tan{k_1 L_1}=\frac{i k_3\tanh{\beta\Delta}-\beta }{\beta\tanh{\beta \Delta}-i k_3 } \label{qnmeig}
\ee 
 with $\Delta=L_2-L_1$. 
The solutions of this equation in the complex $\omega$  plane are displayed in figure \ref{qnmplanefull}. 
 In the graph we can distinguish three kinds of modes\footnote{We thank Cristoforo Iossa for drawing our attention on this distinction}: a finite number of meta-stable long-lived modes, with almost real frequencies and tiny (yet negative) imaginary parts, a finite sequence of ring-down modes, characterized by always decreasing (being negative) imaginary parts with almost constant real part 
 $\omega_R \approx \sqrt{V_2}$ and finally an infinite tower of modes with complex frequencies where both the real and imaginary parts grow approximately `linearly'. These `linearly raising' modes, that we dub `blind' are rather insensitive to the detailed properties of the potential. 

 Finally we noice that poles of the Love number obtained as solutions of  (\ref{metalove}) are metastable modes satisfying the QNM eigenvalue
 equation  (\ref{qnmeig}) with $\beta \Delta $ large and $k_1 L \approx n \pi$. Such modes exists when the cavity is deep enough.
 
Neumann boundary conditions leads to similar results replacing tangent with (minus) cotangent functions. 
 
 \subsection*{Infalling waves}
 
\begin{figure}[t]
\begin{minipage}{0.5\textwidth}
\centering
\includegraphics[width=1\textwidth]{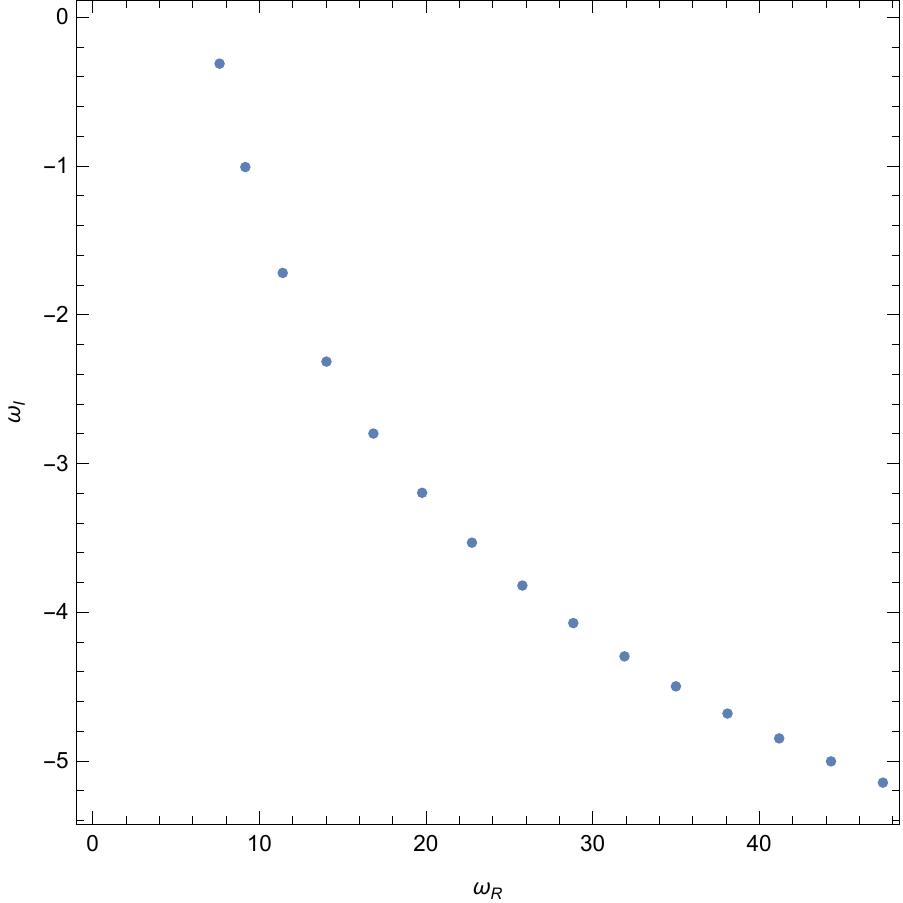}
\end{minipage}
\hspace{0.2cm}
\begin{minipage}{0.4\textwidth}
\centering
\includegraphics[width=1\textwidth]{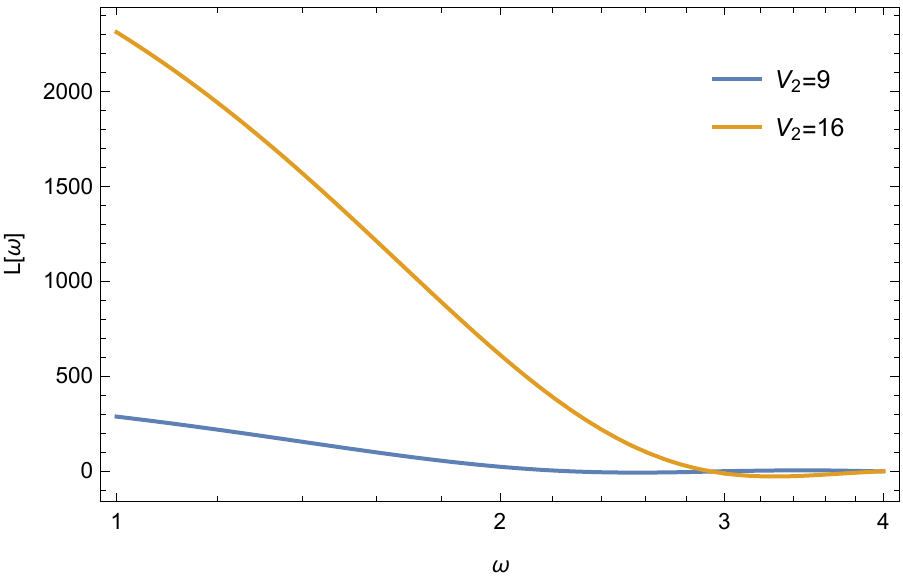}\\
\includegraphics[width=1\textwidth]{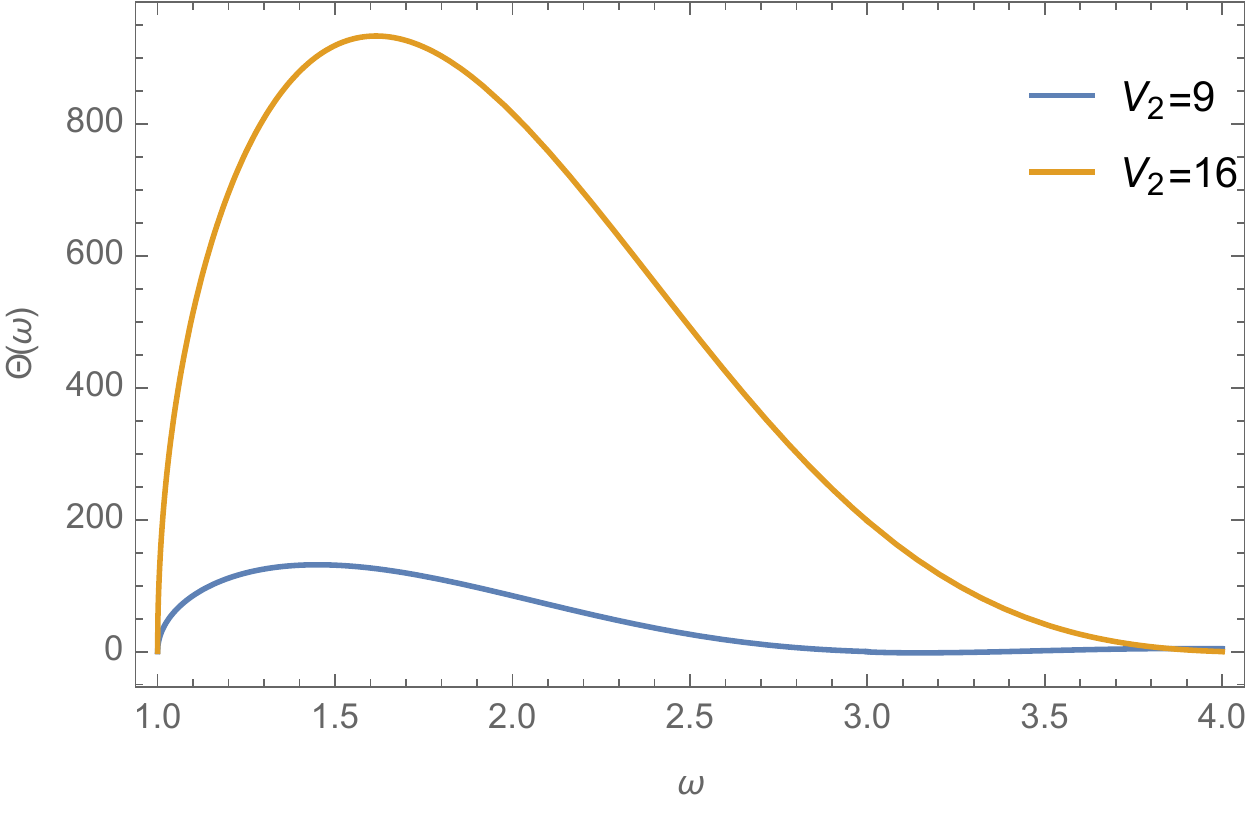}
\end{minipage}
\caption{QNMs and Tidal Love and dissipation numbers for $\ell=0$ waves with Ingoing boundary conditions: a) QNMs for square potential barriers with
$V_1=1$, $V_2=50$, $\Delta=1$, $L_1=1$.  b) Love and dissipation numbers }\label{qnmplanefull2}
\end{figure}

 Incoming waves at a horizon located at $x=0$ are described by a wave function
  \be\label{wavfunc3}
\Psi^{H}(x) =\left\{\begin{array}{lll}
        A_0 e^{-{\rm i} k_1 x}, &~~~~~~~~&   0\leq x\leq L_1\\
        B_0 e^{-\beta x}+   C_0 e^{\beta x}  &&   L_1< x\leq L_2\\
         D_0 e^{i k_3 x}  +\tilde{D}_0 e^{-i k_3 x}  &&   x>L_2
        \end{array} \right.
 \ee
 For the TLNs one  finds
 \be
{\cal L}^{H}_0(\omega)={B_0(\omega)\over C_0(\omega)}= {\b + i k_1 \over \b - i k_1} e^{2 \b L_1}
\ee
 which is generically complex, combining tidal deformations and dissipative effects associated to the `caricature' horizon. Now, there are no poles along the real line. The results for some choices of parameters are shown in figure \ref{qnmplanefull2} (right). 
  The eigenvalue equation for the QNMs become
 \be
\frac{ i \beta}{k_1} =\frac{i k_3\tanh\beta\Delta-\beta}{\beta\tanh\beta \Delta-ik_3 }
\ee 
that admits an infinite tower of complex solutions all with negative imaginary part, see figure \ref{qnmplanefull2} (left).  
 
%
%
%
%
%
%

%
%
%
%

\subsection{QNMs and Tidal Love numbers for $\ell \neq 0$}

For $\ell\neq 0$, the wave function in the three regions can be written as
 \be\label{wavfunc}
\Psi(x) =\left\{\begin{array}{lll}
     A\,  \sqrt{ x } J_\nu (k_1 x) +  \tilde{A}\,  \sqrt{ x } J_{-\nu} (k_1 x)  &~~~~~~~~&   0\leq x\leq L_1\\
        B \, \sqrt{ {-} x }  K_\nu ({-}\beta x) +C \, \sqrt{ x  } K_\nu(\beta x) &&   L_1< x\leq L_2\\
         D\, \sqrt{-  x  } K_\nu ( -{\rm i} k_3  x) + \tilde{D}\,   \sqrt{  x  }  K_\nu ({\rm i} k_3  x) & &   x>L_2
        \end{array} \right.
 \ee
 with $J_\nu$, $I_\nu$ and $K_\nu$ (modified) Bessel functions\footnote{For $\nu=\ell+\ft12$, Bessel functions are related to spherical Bessel functions $J_{\ell+\ft12}(z) = \sqrt{2z\over \pi} j_\ell(z)$.}, $A,\tilde{A},B,C,D,\tilde{D}$ some constants and
 \be
 \nu=\ell+\ft12 
 \ee
  The constant $\tilde{A}$ is determined by the boundary conditions at $x=0$. For example, Dirichlet and Neumann  boundary conditions at $x=0$ ($\Psi(0)=0$ or $\Psi'(x)=0$) corresponds to $\tilde{A}=0$ or $A=0$ respectively. Ingoing boundary conditions mimicking BH horizons correspond to taking both $A$ and  $\tilde{A}$ non vanishing such that $\Psi(x) \sim \sqrt{x} K_{\nu}({\rm i}k_1 x) $  in the first region.     
 
\begin{figure}[t]
\begin{minipage}{0.4\textwidth}
\centering
\includegraphics[width=1\textwidth]{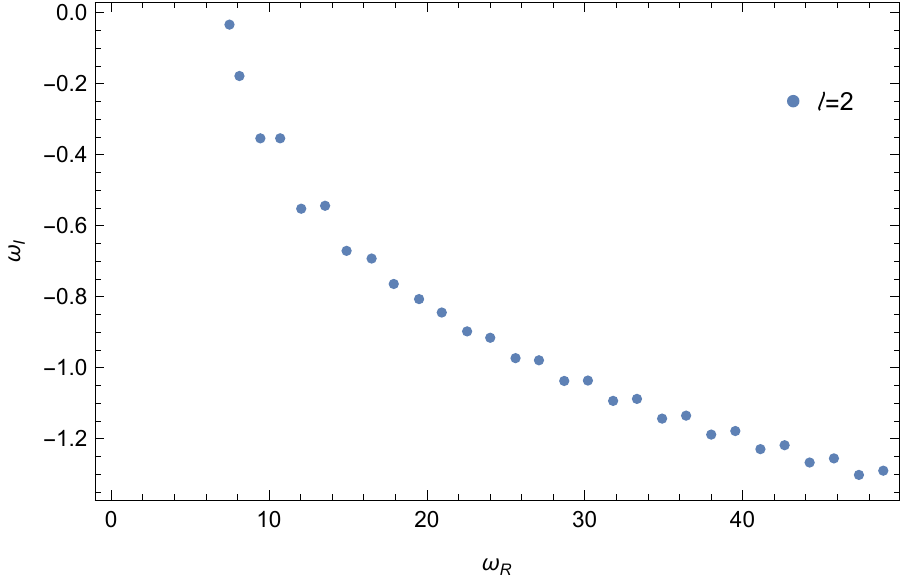}
\end{minipage}
\hspace{0.2cm}
\begin{minipage}{0.5\textwidth}
\centering
\includegraphics[width=1\textwidth]{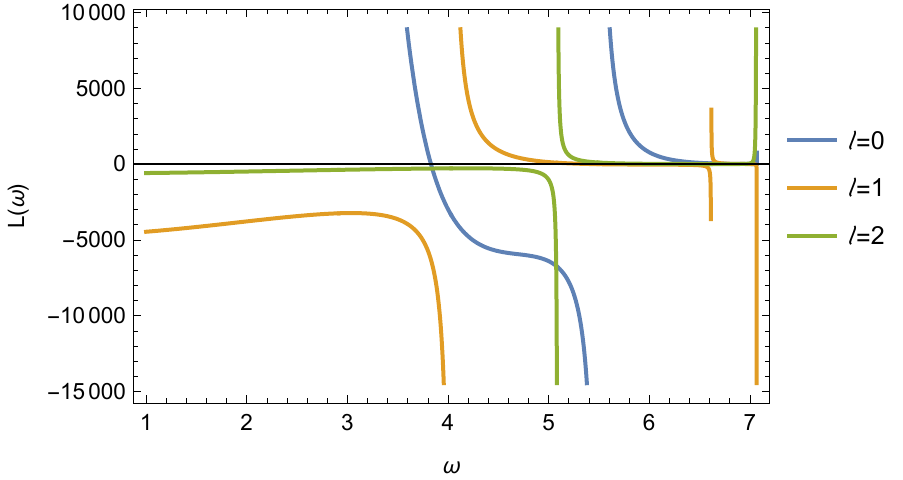}
\end{minipage}
\caption{QNMs and Tidal Love numbers for waves with Dirichlet boundary conditions scattering from square potential barriers with
$V_1=1$, $V_2=50$, $\Delta=1$, $L_1=1$ for various choices of $\ell$. }\label{qnmplanefull3}
\end{figure}

 The remaining constants are then determined  matching the wave function and its first derivative on the left and right sides of the points $x=L_1$ and $x=L_2$. In the near zone (the intermediate region) where $\beta x \ll 1$, the wave function behaves as
\be
{\Psi (x) \over x}  \sim  C\, \Gamma({-}\ft12 {-}\ell)   \, \left( { \beta x \over 2}\right)^{\ell}   +   B\, \Gamma(\ft12+\ell)  \left(- { \beta x \over 2}\right)^{-\ell-1} 
\ee
 The dynamical TLN $\mathcal{L}_\ell(\omega)$  is defined as the ratio between the coefficient of the $x^{-\ell-1}$ (response) and the $x^{\ell}$ (source) terms
 \be
 \mathcal{L}_\ell(\omega)= (-)^{-\ell} { \Gamma(\ft12+\ell) \over \Gamma({-}\ft12{-}\ell)} \left({  \beta \over 2}\right)^{-2\ell-1} \,  {B(\omega) \over C (\omega) }  \ee
For concreteness let us consider the case of scattering in front of an infinite wall, i.e. $\tilde{A}=0$. 
 Proceeding as before one finds for the dynamical Love number
 \be
\mathcal{L}_\ell(\omega) = (-)^{-\ell} { \Gamma(\ft12+\ell) \over \Gamma({-}\ft12{-}\ell)} \left({  \beta \over 2}\right)^{-2\ell-1} \,   \frac{\beta  J_{\nu }( k_1  L_1)
  K'_{\nu }(\beta  L_1)-k_1 J'_{\nu}(k_1 L_1) K_{\nu }(\beta  L_1)}{ \beta  J_{\nu }(k_1  L_1) K'_{\nu }(-\beta L_1)+k_1 J'_{\nu}(k_1  L) K_{\nu }(-\beta L_1)}
 \ee
  Again one finds a real function (no dissipation) with infinite number of poles, see figure  \ref{qnmplanefull3} (right). 
  
On the other hand, QNMs are obtained by setting $\tilde{D}=0$, leading to the eigenvalue equation
\beaq
&& \frac{\beta  J_{\nu }\left(k_1 L_1\right) K'_{\nu}\left(-\beta  L_1\right)+k_1 J'_{\nu}\left(k_1 L_1\right) K_{\nu }\left(-\beta  L_1\right)}{\beta
   J_{\nu }\left(k_1 L_1\right) K'_{\nu}\left(\beta  L_1\right)-k_1 J'_{\nu}\left(k_1 L_1\right) K_{\nu }\left(\beta  L_1\right)}\\
   && \qquad \qquad\qquad \qquad  =\frac{-k_3 K'_{\nu} \left(-i k_3 L_2\right) K_{\nu }\left(-\beta  L_2\right)+i \beta  K_{\nu }\left(-i k_3 L_2\right) K'_{\nu}\left(-\beta  L_2\right)}{- k_3 K'_{\nu}\left(-i
  k_3 L_2\right) K_{\nu }\left(\beta  L_2\right)+{\rm i} \beta  K_{\nu }\left(-i k_3 L_2\right) K'_{\nu}\left(\beta  L_2\right)}\nn
\eeaq
 Again, there is an infinite number of solutions for $\omega$ in the complex plane. The results are displayed in figure \ref{qnmplanefull3} (left). 
 
%

%
%

\section{Geodesic motion on topological star geometries}

In this section, we briefly review top stars and their main properties \cite{Bah:2020pdz, Heidmann:2022ehn} and study the geodetic motion of massless and massive probes. 

\subsection{The five and four-dimensional solution}

Top stars, in their simplest version, are magnetically charged solutions of the Einstein-Maxwell 5-dimensional theory
\be
S_{5} =   \int d^5x \sqrt{g} \left[  {R \over 2 \kappa_5^2} - {1\over 4} F^2  \right]
\ee
 The spacetime is asymptotically $\mathbb{R}^{1, 3} \times S^1_y$.  The metric and gauge field profiles are given by
	\beaq \label{metric}
		ds^2 &=& - f_s(r)dt^2+{dr^2\over f_s(r) f_b(r)}+  r^2(d\q^2+\sin\q^2d\phi^2)+ f_b(r)dy^2\nn \\
		 F_2 &=& P \sin \q d \q \wedge d \phi
		 	\eeaq
with
\be	
	f_s(r)=1-{r_s \over r}\qquad ,\qquad f_b(r)=1-{r_b\over r}\qquad ,\qquad  P^2={3 r_s r_b \over 2 \kappa_5^2}
\ee
The coordinate $y$ is compact $y\sim y+2\pi R_y$. The geometry is regular for $R_y,r_s,r_b$ satisfying the condition
\be
r_s=r_b-{4 r_b^3 \over R_y^2} 
\ee
that implies that $r_s\leq   r_b$ and $R_y \geq 2 r_b$. In order to avoid Gregory-Laflamme instability \cite{Gregory:1993vy} and thermodynamical instability, one has to require \cite{Bah:2021irr} 
\be
r_s < r_b < 2 r_s
\ee
For $r_b>r_s$ the geometry terminates at a cap at $r=r_b$ around which the geometry looks like $\mathbb{R}_{t,y,r}^{2,1}\times S_{\theta,\phi}^2$. For $r_s = r_b$, the geometry near the cap looks like $AdS_3 \times S^2$ sourced by a black string along the $y$ direction. 

After Kaluza-Klein reduction along the compact direction $y$, the top star solution reduces to a four-dimensional (singular) solution of the resulting Einstein-Maxwell-Dilaton theory 
\be
S_4=\int d^4x \sqrt{g} \left[  {R \over 2 \kappa_4^2} -{3\over \kappa_4^2} (\partial \Phi)^2- {e^{-2\Phi}\over 4 e^2} F^2  \right]
\ee
with 
\be
\kappa_4^2 ={\kappa_5^2 \over  2\pi R_y} \qquad , \qquad  e^2={1\over 2\pi R_y}
\ee
 The metric, dilaton and gauge field become
\begin{align} \label{gravity4d}
& ds_4^2 = f_b(r)^{1\over 2} \Bigl [  f_s(r)  dt^2 + \frac{ dr^2}{ f_s(r) f_b(r) } + r^2 (d \q^2 + \sin^2 \q d \phi^2) \Bigr] \\ 
& e^{2 \Phi} =  f_b(r)^{- {1\over 2} } \\
&  F = e\, \tilde P \sin \q d \q d\phi
\end{align}
The four dimensional  ADM mass and magnetic charges are
\be \label{ADM mass}
\mathcal{M} = \frac{2 \pi}{\k_4^2} (2 r_s + r_b) \qquad, \qquad \tilde{P}^2=  {3 r_b r_s\over 2 \kappa_4^2 } 
\ee
The 4-dimensional solutions can then be classified according to the values of $\mathcal{M}$ and $\tilde{P}$. Inverting \eqref{ADM mass}, one finds two types of solutions
  \begin{align}
	&  r^{I}_s = \frac{\k_4^2}{8 \pi } \left( \mathcal{M} + \sqrt{  \mathcal{M}^2-\ft{64\pi^2 \tilde{P}^2 }{ 3 \kappa_4^2}  } \right) 
	\qquad ~, \qquad  r^{I}_b=  \frac{\k_4^2}{4 \pi } \left( \mathcal{M} - \sqrt{  \mathcal{M}^2-\ft{64\pi^2 \tilde{P}^2 }{ 3 \kappa_4^2}  } \right)\\
	& r^{II}_s = \frac{\k_4^2}{8 \pi } \left( \mathcal{M} - \sqrt{  \mathcal{M}^2-\ft{64\pi^2 \tilde{P}^2 }{ 3 \kappa_4^2}  } \right) 
	\qquad , \qquad  r^{II}_b=  \frac{\k_4^2}{4 \pi } \left( \mathcal{M} + \sqrt{  \mathcal{M}^2-\ft{64\pi^2 \tilde{P}^2 }{ 3 \kappa_4^2}  } \right)
		\end{align}
Requiring that $r_b>r_s$ and that both be real one finds that the two solutions exists in the ranges
\begin{align}
	I: \qquad  {1\over \sqrt{6} } \leq  {2 \pi | \tilde P|  \over {\cal M} \kappa_4} \leq  {\sqrt{3} \over 4}  \qquad , \qquad 
	II : \qquad    {2 \pi | \tilde P|  \over {\cal M} \kappa_4} \leq  {\sqrt{3} \over 4}  
		\end{align}
We stress that contrary to the over-rotating JMaRT solutions \cite{Jejjala:2005yu}, the top stars solutions admit a regime of parameters such that both a black object and a smooth horizonless solution have the same asymptotic properties.  However, stable topological stars do not coexist with the black string solution\footnote{We thank Pierre Heidmann for a clarifying comment in this respect.}.

\subsection{Light rings}

In the Hamiltonian formalism, the geodetic motion of a particle of, possibly zero,  mass $\mu$ moving in a metric $g_{\mu\nu}$  is governed by the Hamiltonian
\be
\mathcal{H}=\ft12 g^{\mu \nu}P_\mu P_\nu=-{\mu^2\over 2}   \label{ham}
\ee
 with $P_\mu$ the canonical conjugate momenta, so that $\dot{x}^\mu = \partial\mathcal{H}/\partial{P}_\mu$. For top stars in $D=5$, one finds    
\begin{align}\label{canmom}
P_t&=-E=- f_s(r)\dot{t}\quad,\quad P_r={\dot{r}\over f_s(r) f_b(r)}\nn\\ 
P_\theta &=r^2 \dot{\theta}\quad,\quad P_\phi=J_\phi =r^2\sin^2\q \dot{\phi}\quad,\quad P_y=f_b(r)\dot{y}
\end{align}
with the momenta $E$, $J $ and $P_y$ conserved thanks to the large isometry of the background.  The radial and angular dependence in 
\be
2\mathcal{H}+\mu^2=-{E^2\over  f_s(r) }+{P_y^2\over f_b(r) }+P_r^2 f_b(r) f_s(r)+\mu^2 +{1\over r^2} \left( P_\q^2+{J_\phi^2\over  \sin^2\q} \right) =0
\ee
 can be separated and solved by 
\begin{align}\label{P2}
P_r^2  &= Q_{r,{\rm geo}} (r)={ r^2 \left[ E^2\, f_b(r)-P_y^2  f_s(r) \right]  -(K^2+\mu^2 r^2) f_s(r)f_b(r)\over  r^2f_b^2(r) \, f_s^2(r)}\\
P_\theta^2 &=Q_{\theta,{\rm geo}} (\theta)=  K^2-{J_\phi^2\over \sin^2\q} \label{Qgeo}
\end{align}
with $K^2\ge 0$ a separation constant representing the total angular momentum of the probe.  

Thanks to spherical symmetry, without loss of generality, one can set $\theta = \pi/2$ and thus $\dot\theta=0=P_\theta$ so that 
$K=|J_\phi|$. Then, one can take $J_\phi = r^2 \dot\phi = K>0$. For instance, the deflection angle of a probe that reaches a turning point at $r_*$ is given by
\be
\Delta \phi =2\int_{\infty}^{r_* }  {\dot \phi\over \dot{r} } dr  -\pi   =2 K  \int_{\infty}^r  { dr \over  r^2 f_s(r) f_b(r)\, P_r(r)}  -\pi 
\ee
with $P_r$ given by (\ref{P2}), that can be written in terms of elliptic integrals. 

A simpler problem, that is more interesting for our present purposes, consists in finding critical geodesics that form the so called light-ring or attractor points of massive motion, obtained at fixed points (zero velocity and acceleration) of the radial motion. The defining conditions read
\be \label{critical}
Q_{r_{\rm geo}} (E_c,r_c)=Q_{r,{\rm geo}}'(E_c, r_c)=0
\ee 
with the derivative performed w.r.t. $r$.
For top stars one can write
\be
 Q_{\rm geo}(r)={ N(E, r)  \over r^4 f_b^2(r) \, f_s^2(r)}  
\ee
with
\be
N(E, r) =  r^3 \left[ E^2\, (r{-}r_b) -P_y^2  (r{-}r_s) \right]  -  (K^2+\mu^2 r^2)  (r{-}r_b)(r{-}r_s) \label{ner}
\ee
a quartic polynomial. 
The critical equations $N(E_c, r_c) =N'(E_c,r_c)=0$  become
\beaq
E_c^2 & =&  {(r_c{-}r_s)^2  \left[ K^2(3 r_b{-}2 r_c) {+}\mu^2 r_b r_c^2 \right] \over r_c^4 ( r_b{-}r_s) } \nn\\
 P_y^2 &=&   {(r_c{-}r_b)^2 \left[  K^2(3 r_s{-}2 r_c) {+}\mu^2 r_s r_c^2  \right] \over r_c^4 ( r_b{-}r_s) }    \label{pyc}
\eeaq
The solution for $r_c$ is obtained from the second equation as a root of a quartic polynomial.
Solutions exists when both right hand sides in (\ref{pyc}) are positive and $r_c\ge r_b$.

The case  $P_y=0$ is simpler. The  second equation in (\ref{pyc})  leads to a cubic or quartic polynomial equation with roots
\beaq \label{critradii}
 \mu = 0: && \qquad r_{c,0}  = r_b  \quad , \quad  
r_{c,1} =  \ft32 r_s\nn\\
\mu \neq 0: && \qquad r_{c,0}  = r_b  \quad , \quad  
r_{c,\pm} =  {K^2\pm K \sqrt{K^2-3 \, r_s^2\, \mu^2 } \over r_s \, \mu^2}  
\eeaq
  
The solutions $r_{c,-}$ and $r_{c,+}$ correspond to the unstable photon ring and the ISCO (Inner-most Stable Circular Orbit) respectively.
In particular the ISCO is present only for $\mu\neq 0$ and for sufficiently large $K$: $K^2>3r_s^2 \mu^2$. For small $\mu$ the ISCO is widely separated from the unstable external photon-ring  
 \beaq
r_{c,+} & =& r_{\rm ISCO} \approx  {2K^2\over r_s\mu^2}  - {3\over 2}r_s + \ldots \\  
r_{c,-} & = & r_{\rm photon-sphere} \approx  {3\over 2}r_s - {9 r_s^3\mu^2 \over  8K^2}  + \ldots 
\eeaq 
so in this limit $r_{\rm ISCO} \gg r_{\rm photon-sphere}$. In this respect a top star could be surrounded by an accretion disk very much as a putative BH. 

For a given $r_c$ in the allowed range, where both RHSs of (\ref{pyc}) are positive, the corresponding critical energies are given by (\ref{pyc}). 

Notice that at $r=r_{c,0}  = r_b$, the momentum is non-vanishing $Q_{r_{\rm geo}}'(E_c,r_c) \neq 0$. Still this point represents a fixed point of the motion where $\dot{r}=\ddot{r}=0$. The physical explanation of this phenomenon is the infinite `effective' mass $m_{\rm eff}(r) = P_r/\dot{r} = 1/f_bf_s$ of the probe at $r=r_b$ where $f_b(r_b)=0$.

\subsection{Lyapunov exponent}
The motion close to unstable light rings is characterized by a chaotic behavior, quantified by the Lyapunov exponent $\lambda$ \cite{Cardoso:2008bp, Maldacena:2015waa, Bianchi:2020des}.
 The Lyapunov exponent parametrizes the vanishing rate of the velocity of a geodesic approaching the light ring at $r= r_c$ at critical energy $E_c$ \cite{Maldacena:2015waa, Bianchi:2020des}
\be \label{drdt}
{d r \over d t} \approx -2 \lambda (r-r_c)  
\ee
Integrating this equation, one finds that the time delay $\Delta t$ experimented by a geodesics starting at small distance $\Delta r_0$ from the photon sphere  to reach a point at distance  $\Delta r$ 
and getting back is given by\footnote{This definition motivates the extra factor 2 with respect to \cite{Cardoso:2008bp}.}
\be
\Delta t =2 \int_{r_c+\Delta r}^{r_c+\Delta r_0} { dt \over dr  } dr  \approx   {1 \over \lambda}  \log {\Delta r \over \Delta r_0}  
\quad \Rightarrow \quad   \Delta r \approx e^{\lambda \Delta t}
\ee
Making  use of (\ref{canmom}) one can write
\be \label{drdt1}
{d r \over d t} = {\dot{r} \over \dot{t}} =- { {f_s (r)}^2 f_b(r) \sqrt{Q_{\rm geo} (r) } \over E_c }  
\ee
leading to
\be\label{Lyapunov}
\lambda  = { {f_s (r_c)}^2 f_b(r_c) \sqrt{Q''_{\rm geo}(r_c) } \over  2 \sqrt{2} E_c   }   =  {(r_c - r_s) \sqrt{N''(E_c, r_c)} \over 2 \sqrt{2} E_c r_c^3 }
\ee
with $N(E, r)$ given by (\ref{ner}). 

In the case of BHs, QNM frequencies of prompt ring down modes can be estimated in a semiclassical approximation by the WKB formula (\ref{wwkb})  
\be  
\omega^{\rm geo}_n  \approx  E_c(r_c) -{\rm i} (2n+1) \lambda  
\ee   
where $n$ is called the overtone number. This formula provides a good estimate of QNM frequencies for geometries where metastable and ring-down modes are well separated\footnote{We thank Pierre Heidmann for pointing out this subtlety.}. We will later use this formula as seed for cases where no stable light-ring is present.

\section{Quasi-normal modes}  
   
\subsection{The wave equation}

We now pass to consider the linearized scalar pertubations around top star geometries that is described by the wave equation
\be
(\Box-\mu^2)\Phi=0
\ee
This equation can be separated into two ordinary differential equations describing the radial and angular motions via the ansatz
\be
\Phi=\exp\left(-i \omega t+i p y+i m_\phi \phi \right)R(r)S(\q)
\ee
The angular equation boils down to the standard equation for spherical harmonics in four dimensions: 
\be
\Big[{1\over \sin\q}\partial_\q(\sin\q \partial_\q )-{m_\phi^2\over \sin^2\q}\Big]S(\q)=-\ell(\ell+1)S(\q)
\ee
On the other hand the radial equation  can be written as
\be
(r{-}r_s)(r{-}r_b) R''(r){+} (2r{-}r_s{-}r_b) R'(r){+} \left( {\omega^2 r^3 \over r{-}r_s} {-} {p^2 r^3 \over r{-}r_b}{-}K^2{-}\mu^2 r^2 \right) R(r){=}0 \label{eqr}
\ee
 The equation can be put in the canonical form 
\be
\Psi''(r)+Q(r)\Psi(r)=0 \qquad , \qquad R(r)={\psi(r)\over r\sqrt{f_b(r)f_s(r)}} 
\ee
with
\be
Q(r) ={ 4 r^3 \left[ \omega^2 (r-r_b) -p^2 (r-r_s)\right] -4 \left[ \ell(\ell+1)+\mu^2 r^2 \right] (r-r_b)(r-r_s)+(r_b-r_s)^2 \over 4 (r-r_b)^2  (r-r_s)^2}   \label{qwave}
\ee
It is easy to check that $Q(r)$ coincide with $Q_{\rm geo}$ given in (\ref{Qgeo}) after discarding the last term in the numerator in (\ref{qwave}) and identifying $K^2=\ell(\ell+1)$.  
Light rings are determined by the critical equations
\be
Q(r_c) =Q'(r_c) =0
\ee

\begin{figure}[t]
\centerline{ 
 \includegraphics[width=0.3\textwidth]{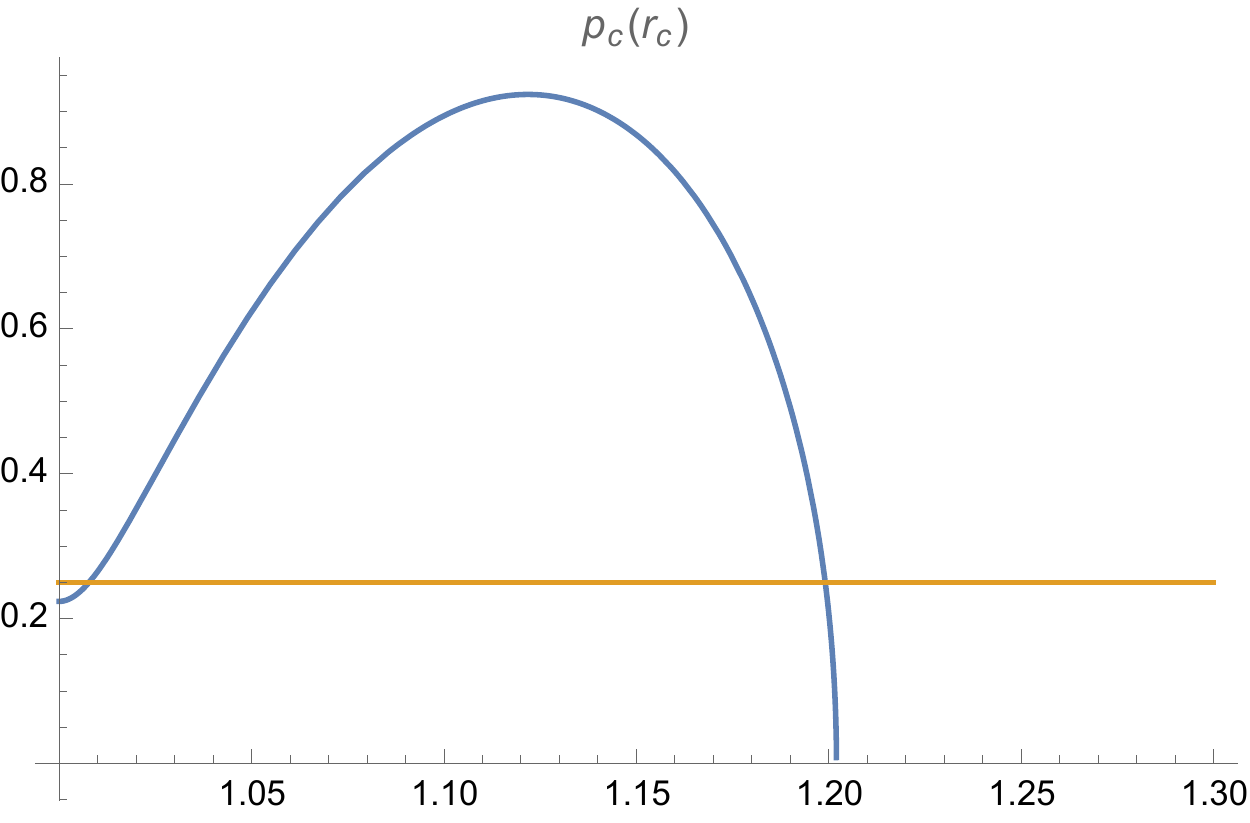}\quad \includegraphics[width=0.3\textwidth]{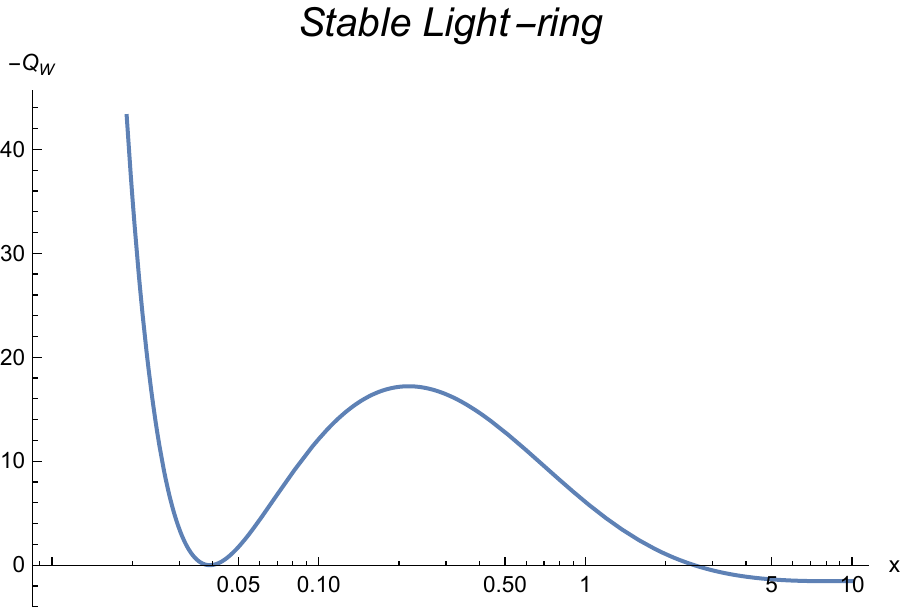}\quad 
 \includegraphics[width=0.3\textwidth]{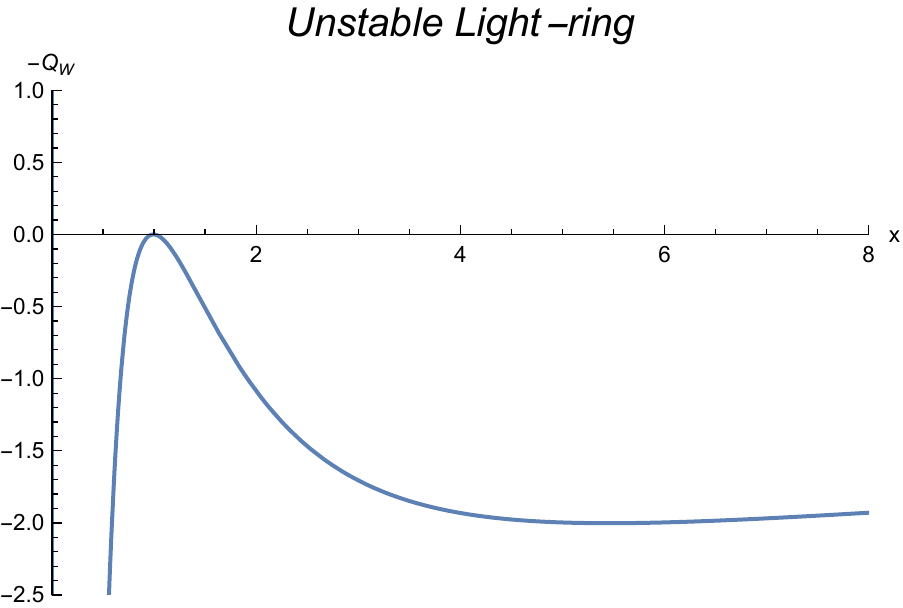}
}
\caption{ Light rings: a) Dependence of the y-momentum $p$ on the critical radius for a top star with $r_s=0.8$, $r_b=1$, $\mu=0$, $\ell=2$.  
For $p=0.25$ there are two light rings associated to the two intersections. b) Potential for $\omega=\omega_c(r_{c1})$.
c) Potential for $\omega=\omega_c(r_{c2})$.
  }\label{figQs}
\end{figure}

 As before, we look for solutions in $p$ and $\omega$
\begin{align}
\omega_c(r_c)^2 &= \frac{\left(4 r_c-3 r_s\right) \left(r_b-r_s\right){}^2-4 K^2\left(r_c-r_s\right){}^2 \left(2 r_c-3 r_b -r_c^2  r_b \mu^2\right)
}{4 r_c^4 \left(r_b-r_s\right)}\nn\\
p_c(r_c)^2& = \frac{\left(4 r_c-3 r_b\right) \left(r_b-r_s\right){}^2-4 K^2 \left(r_c-r_b\right){}^2
\left(2 r_c-3 r_s -r_c^2  r_s \mu^2\right)}{4 r_c^4 \left(r_b-r_s\right)} 
\end{align}
In figure \ref{figQs} we plot the function $p_c(r_c)$ for given values of $\ell$, $r_s$, $r_b$ and $\mu$. Given a momentum $p$, a light ring exists if the line $f(r_c)=p$ intersects the curve $p_c(r_c)$ at some point. For example, for the choice of parameters in \ref{figQs} there are two light rings, an internal stable one, and an external unstable one.

The plots for $-Q(r)$ corresponding to the two critical values $\omega_c(r_{c,1})$, $\omega_c(r_{c,2})$  are displayed in figure \ref{figQs}.   On the other hand for $p<p_c(r_b)$ there is a single unstable light ring.  Finally  above certain value of $p$ there is no light-ring at all.  
For small $p$ we expect then only prompt ring down QNMs, while for $p$ in the intermediate region we expect both metastable and ring down QNMs.

 \subsection{WKB analysis}
 
 QNMs can be determined by a WKB approximation of the solution. Barring truly stable modes, related to stable orbits, we should distinguish between unstable and (meta)stable modes, corresponding to frequencies $\omega$ near a maximum or a minimum respectively of the `potential' $-Q(x)$.  We will refer to the former as prompt ring down modes, bearing in mind the analogy with BH mergers.  
Prompt ring down frequencies can be estimated using  Bohr-Sommerfeld quantization condition\footnote{The nomenclature $a_D$ and $a$ for the `periods' anticipates the analysis in the context of (quantum) Seiberg-Witten curves.} 
   \be
 a_D={1\over \pi} \int_{r_2}^{r_3} \sqrt{ Q(\omega_n^{\rm prmpt}, r)}\,  dr =n+\ft12  \label{adquant}
 \ee
  with the integral performed in the interval between two zeros $r_2,r_3$ where  $Q(r)<0$. 
 The integral can be approximated by assuming that $\omega$ is almost the critical one plus a small imaginary part. To linear order in the imaginary part, one finds 
\be
\omega_n^{\rm prmpt}\approx  \omega_c - {\rm i} \, { (n+\ft12) \sqrt{ 2 Q''(x_c) } \over \partial_w Q(x_c) } \label{wwkb}
\ee
with $\omega_c,x_c$ obtained by solving the critical equations (\ref{lightringeq}). This previous expression coincides with \eqref{Lyapunov} by replacing $Q-$function for geodesics with the same for waves equation. The results will be later used as seeds for scan QNMs frequencies based on q-SW periods (see Table \ref{tabQNMp}).

 Metastable states instead, live near the local minima of the potential $-Q(x)$ and are related to quantization of the $a$-cycle. The real part  of the frequencies  can be estimated by the WKB formula
 \be
 a={1\over \pi} \int_{r_1}^{r_2} \sqrt{ Q(\omega_R^{\rm bnd}, r)}\,  dr =n+\ft12  
 \ee
where now the integral runs along the classically allowed region $[r_1,r_2]$ where $Q(x)>0$. The imaginary part can be estimated as the inverse of the life-time, {\it i.e} the time the particle takes to bounce $N=e^\Gamma$ times ($e^{-\Gamma}$ being the probability of getting through the barrier)  between the walls before escape \cite{Gasiorowicz:104258}
 \be
 \omega_I^{\rm bnd} = - {1\over \tau} =-{1\over T} e^{-\Gamma} \approx  -{\omega_R^{\rm bnd}\over 2\pi  } e^{-2 \int_{r_2}^{r_3} \sqrt{-Q(\omega_R^{\rm bnd}, x)}\,  dx}
 \ee 
  and it is exponentially suppressed with the length of the dual $a_D$ cycle.

\subsection{ Numerical computation of QNMs}

In this section we apply numerical methods to the computation of QNM frequencies. 
Here by a QNM, we mean a solution $R(r)$ of (\ref{eqr}) regular at $r=r_b$ satisfying outgoing boundary conditions at infinity
  \be
  R(r) \underset{r\to r_b}{\sim} 0\quad\quad , \quad\quad  R(r) \underset{r\to \infty}{\sim} e^{{\rm i} r  \tilde \omega   }  
    \ee
   with
\be
\tilde \omega= \sqrt{\omega^2-p^2-\mu^2}
\ee 
 We will consider two numerical methods: Direct Integration and Leaver, and show perfect agreement between the two.

\subsubsection{Direct Integration}

\begin{figure}[t] 
\centering
\includegraphics[width=0.45\textwidth]{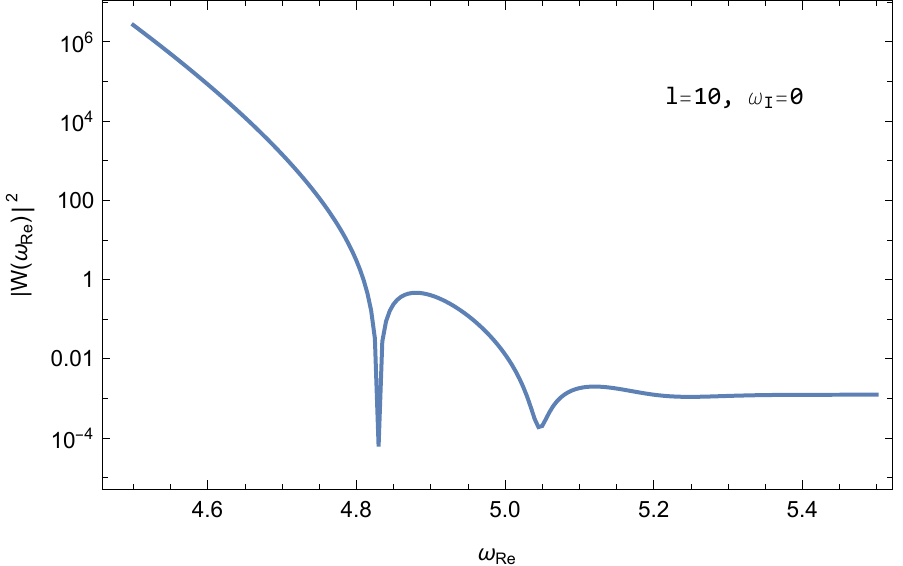}
\quad  \includegraphics[width=0.45\textwidth]{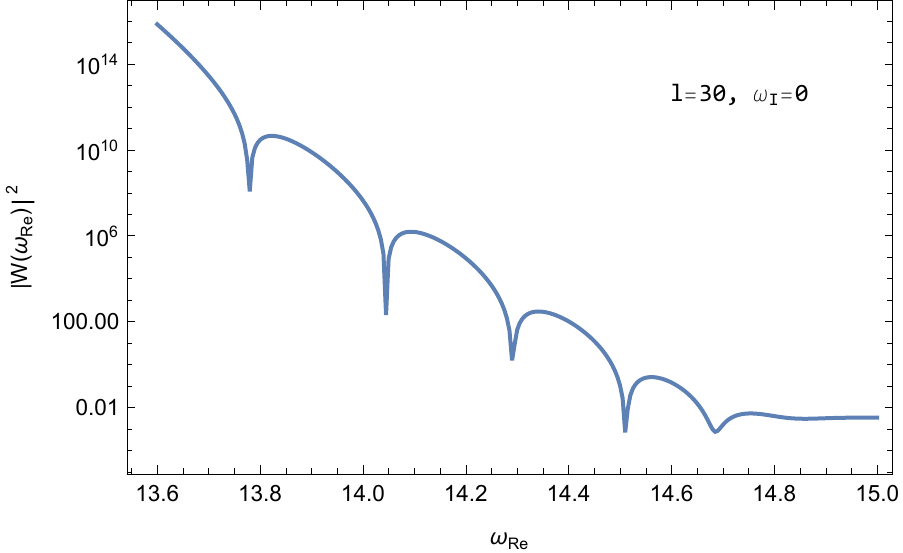}
\caption{ $r_b=1$, $r_s=0.8$, $p=\mu=0$ for $\ell=10$ and $\ell=30$. }
\label{graphbw}
\end{figure}

\begin{figure}[t] 
\centering
\begin{minipage}{0.45\textwidth}
\includegraphics[width=1\textwidth]{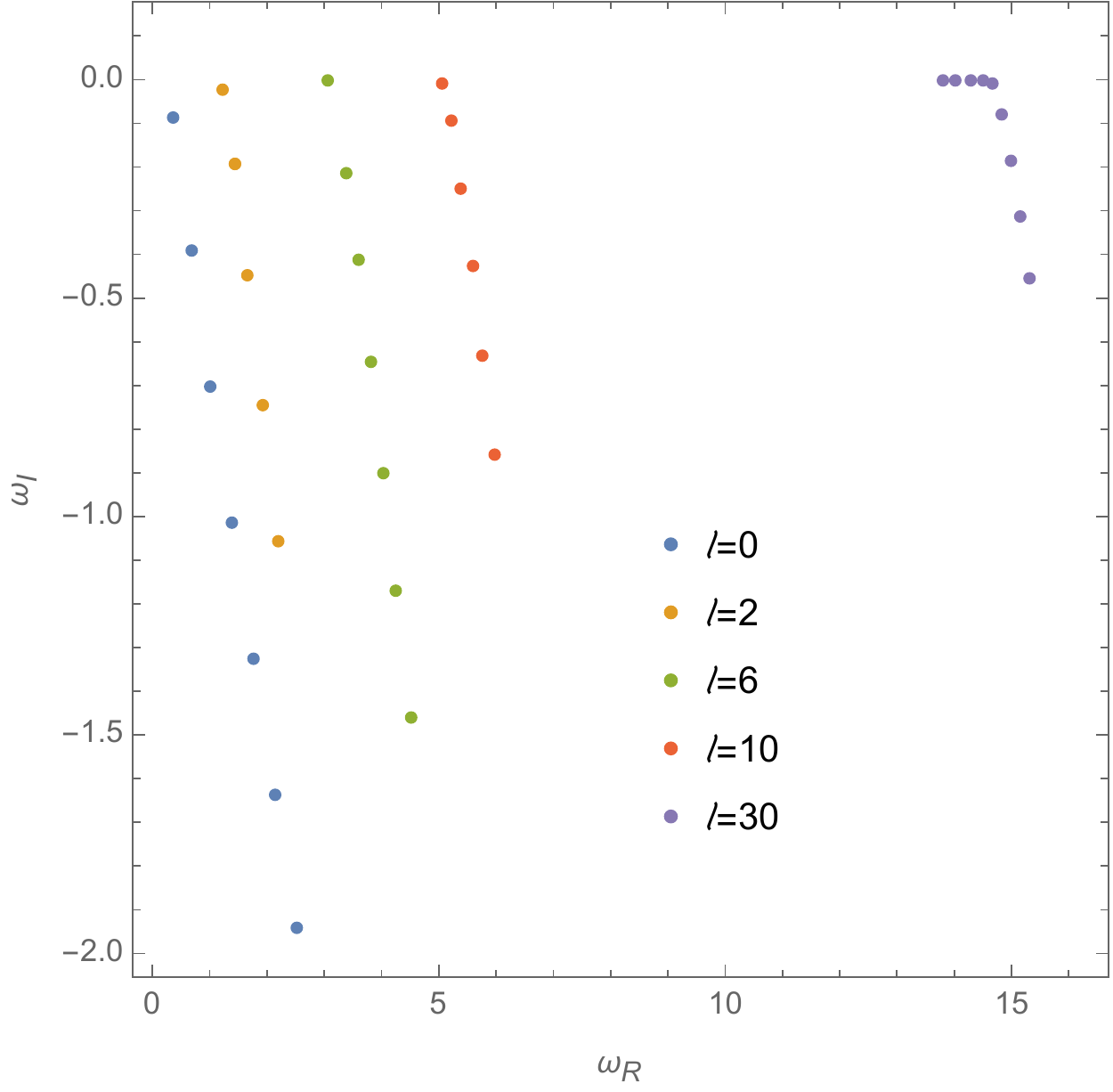}
\end{minipage}
\begin{minipage}{0.45\textwidth}
\includegraphics[width=1\textwidth]{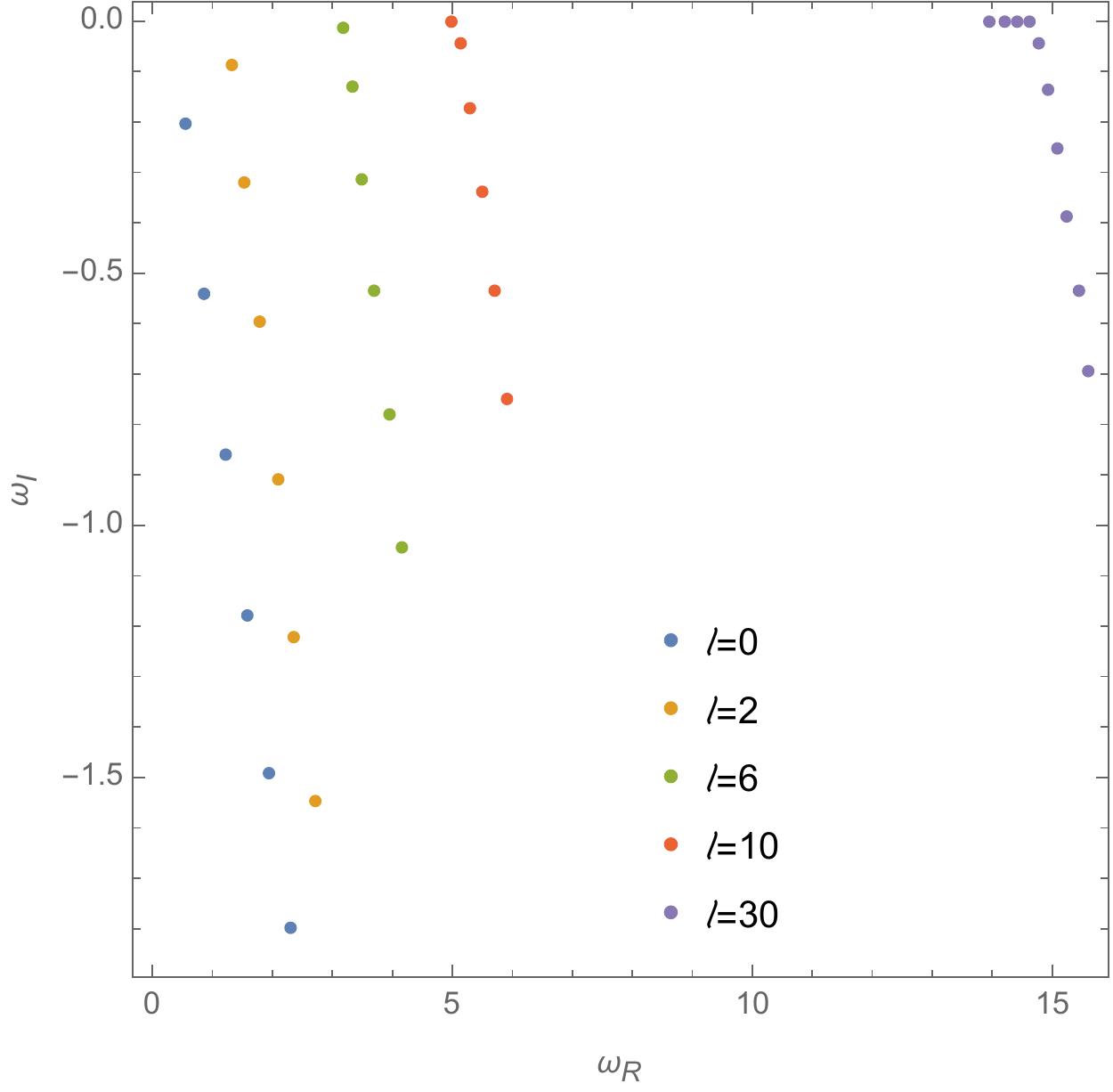}
\end{minipage}
\caption{ QNMs for top stars with $r_b=1$, $r_s=0.8$, a)$p=0$ b) $p=0.25$. }
\label{graphbw1}
\end{figure}

 Let us start by a direct integration of the differential equation with prescribed boundary conditions. 
 A QNM follows the ansatz  
  \beaq
 R_{\rm cap 0 }(r)  & \underset{r\to r_b}{\approx} &    \sum_{n=0}^N  a_n\,  (r-r_b)^{  n+\lambda_{\rm cap}  }   \nn\\
 R_{\rm inf 0}(r)  & \underset{r\to \infty}{\approx} &      e^{{\rm i} \tilde{\omega} r \ } r^{ {-}1+\lambda_{\rm inf} }  \,  \sum_{n=0}^N {c_n\over r^n} 
  \eeaq
 with $N$ a positive integer\footnote{In this paper we take $N=3$.} , $a_n$,  $c_n$ some coefficients and
 \be
   \lambda_{\rm inf} ={p^2(2r_b{+}r_s) {+}\mu^2(r_b{+}r_s){-}\omega^2 (r_b{+}2 r_s)  \over  2 {\rm i} \tilde{\omega} }\qquad , \qquad 
    \lambda_{\rm cap} ={ |p| r_b^{3/2}\over \sqrt{r_b-r_s} }  
 \ee
 The coefficients $a_n$ and $c_n$ are determined recursively starting from $a_0$, $c_0$ from the differential equation (\ref{eqr}). Since the Wronskian does not depend on the overall normalization of the solutions, we can set $a_0=c_0=1$.
 We denote by $R_{\rm cap}(r)$ a numerical solution of the differential equation, matching the finite sum $R_{\rm cap0}$ (and its first derivative) at a point slightly bigger than $r_b$. Similarly $R_{\rm inf}(r)$ is obtained by solving  the differential equation with boundary conditions given by the value of $R_{\rm inf0}$ (and its derivative) at a point $r_\infty>>r_b$.  
 
   The QNM frequencies are the zeros of the Wronskian of the two solutions
   \be
   W_{\rm in,out} (\omega)=R_{\rm cap}(r_e) R'_{\rm inf}(r_e) -R'_{\rm cap}(r_e) R_{\rm inf}(r_e)=0
   \ee
   with $r_e$ an arbitrary point in between\footnote{We notice that solutions of this equation are independent of the choice of the point $r_e$}. This equations be solved numerically, scanning for solutions in the complex $\omega$-plane. 
 
     In figures \ref{graphbw} (right) we display the function $W_{\rm in,out}(\omega)$ for some specific choices of the top star parameters with $\omega$ real. 
 The minima along the real line are used as seeds for a scanning of zeros of $W_{\rm in,out}(\omega)
 $ in the complex $\omega$-plane. The results 
 are displayed in \ref{graphbw} (left). We find again a pattern of metastable, prompts and `blind' modes.

Let us notice that all the QNMs we found have negative, sometimes very small, imaginary parts. This means that at least at the linearized level top stars are stable w.r.t. massive scalar perturbations, including those with KK momentum along the circle $S^1_y$. In the following we will refine our analysis and address the issue of tidal deformability relying on qSW techniques.

%
%
%

\begin{table}[]
\centering
\begin{tabular}{|c|c|c|}
\hline
$\ell$ & WKB     & Direct integration       \\ \hline
6      & $3.10$ & $3.16 -  \rm{i} \ 0.014$       \\ \hline
10     & $4.90$   & $4.97 -  \rm{i} \ 0.0007 $    \\ \hline
30     & $13.85$ & $13.93- \rm{i} \ 7.06 \times 10^{-14} $ \\ \cline{2-3} 
       & $14.11$ & $14.19 -  \rm{i} \ 1.23 \times 10^{-9} $  \\ \cline{2-3} 
       & $14.35$ & $14.42 -  \rm{i} \ 3.64 \times 10^{-6} $  \\ \cline{2-3} 
       & $14.56$ & $14.62 -  \rm{i} \ 0.0018 $      \\ \hline
\end{tabular}
\quad
\begin{tabular}{|c|c|c|}
\hline
$\ell$ & WKB     & Direct integration       \\ \hline
8      & $4.05$ & $4.09 - \rm{i} \ 0.006 $      \\ \hline
10     & $4.95$   & $4.99 -  \rm{i} \ 0.0013$    \\ \hline
30     & $13.90$ & $13.96 -  \rm{i} \ 2.06 \times 10^{-13} $ \\ \cline{2-3} 
       & $14.17$ & $14.22 -  \rm{i} \ 3.04 \times 10^{-9} $  \\ \cline{2-3} 
       & $14.40$ & $14.45 - \rm{i} \ 7.76 \times 10^{-6} $  \\ \cline{2-3} 
       & $14.60$ & $14.64 - \rm{i} \ 0.003 $      \\ \hline
\end{tabular}
\caption{Metastable modes $r_b$=1, $r_s=0.8$, left)  $p=0.25$, $\mu=0$, right) $p=0.3$, $\mu=0.1$. }
\end{table}

\subsubsection{Leaver method}

The Leaver method \cite{Leaver:1985ax,Leaver:1990zz} has been recently applied in \cite{Heidmann:2023ojf} to the study of QNMs for top stars with $p=\mu=0$. Here we implement the Leaver algorithm for the general case of  waves with non-zero mass and/or KK momenta. The results will be compared against those obtained by Direct Integration and Seiberg-Witten techniques. 
 
  We start from the ansatz
\be
R(r)=e^{i\tilde{\omega}r}(r-r_b)^{\lambda_b}(r-r_s)^{\lambda_s}\sum_{n=0}^\infty c_n\left({r-r_b\over r-r_s}\right)^n \label{ansatz}
\ee
with  
\be
\lambda_b={p r_b^3/2\over \sqrt{r_b-r_s}}\quad,\quad \lambda_s=-1-{p r_b^{3/2}\over \sqrt{r_b-r_s}}-{\rm i} {p^2(2r_b+r_s)+r_s(\mu^2-2\omega^2)+r_b(\mu^2-\omega^2)]\over \tilde{\omega} }
\ee
The ansatz (\ref{ansatz}) reproduces the right asymptotic behaviour of $R(z)$  at the cap $r\approx r_b$ and infinity $r >> r_b$
and leads to  three term recursion relation:
\begin{align}
&\a_n c_{n+1}+\b_n c_n+\g_n c_{n-1}=0,\quad n=0,1,2,...
\end{align}
with $c_{-1}=0$, and $c_0=1$. The  coefficients $\alpha_n$, $\b_n$, $\g_n$ carry the information about the gravity background, the frequency and momenta of the wave, and are given by:
\begin{align}
\a_n&=(n+1)\left(n+1+{2p r_b^{3/2}\over\sqrt{r_b-r_s}}\right)\\\nn
\b_n &= \left( 2 n+1  + {2 p r_b^{3/2}\over \sqrt{r_b-r_s} } \right) \left[- n+ \lambda _s+\ft12 +i \tilde\omega\left(r_b-r_s\right) \right] -l(l+1)-\ft{1}{2}\nn\\
&\qquad +\frac{p^2 r_b^2 \left(3 r_s-2  r_b\right)}{r_b-r_s}-\mu ^2 r_b^2+\frac{\omega ^2 r_b^3}{r_b-r_s}\nn\\
\g_n&=(n-1-\lambda_s)^2-{r_s^3\omega^2\over r_b-r_s}
\end{align}
Quasi-normal frequencies are determined by the continuous fraction equation:
\be
\b_n={\a_{n-1}\g_n\over\b_{n-1}-{\a_{n-2}\g_{n-1}\over \b_{n-2}-...}}+{\a_{n}\g_{n+1}\over\b_{n+1}-{\a_{n+1}\g_{n+2}\over \b_{n+2}-...}}
\ee
 In the Appendix we compare the results obtained by this method against those resulting from a direct integration of the differential equation.

 \section{Quantum Seiberg-Witten curves for top stars}

In this section we study the wave dynamics using SW techniques \cite{Seiberg:1994aj, Seiberg:1994rs, Nekrasov:2009rc}. We compute QNM frequencies and TLNs that characterize the response of top stars to tidal deformations.

\subsection{Gauge gravity dictionary}

The dynamics of ${\cal N}=2$ supersymmetric SU(2) gauge theory with $(N_L,N_R)=(1,2)$ fundamentals 
living on a Nekrasov-Shatashvilli $\Omega$-background with $\epsilon_1=\hbar=1$, $\epsilon_2=0$ \cite{Nekrasov:2009rc},
is described by the differential equation \cite{Bianchi:2021mft}
\be
\left[ q z^2 P_L(z \partial_z+\ft{1}{2} ) +z P_0(z \partial_z ) + P_R(z \partial_z -\ft{1}{2}) \right] f(z)=0 \label{diffq}
\ee
with  
\be
P_L(x)=(x{-}m_3) \quad, \quad P_0(x)=x^2{-}u{+}q(x{+}\ft12{-}m_1{-}m_2{-}m_3) \quad, \quad P_R(x)=(x{-}m_1)(x{-}m_2) 
\ee
Writing
\be
f(z) =e^{-{q z \over 2}} z^{{m_1+m_2\over 2}}  (1+z)^{-{1+m_1+m_2\over 2} } \Psi(z)
\ee
equation (\ref{diffq}) takes the Schr\"odinger-like form
\be
\Psi''(z) +Q(z) \Psi(z) =0 \label{eqc}
\ee
with
\be\label{swform}
Q(z) =-{q^2\over 4}{+}{1{-}(m_1{-}m_2)^2\over 4z^2}{+}{1{-}(m_1{+}m_2)^2\over 4(1+z)^2}{-}{m_3 q\over z}{+}{1{-}2(m_1^2{+}m_2^2){+}2q(m_1{+}m_2{-}1){+}4u\over 4z(1+z)}
\ee
 Equation (\ref{eqc}) with $Q$ given by  (\ref{swform}) corresponds to a Confluent Heun Equation (CHE) in its normal form. 
  The same equation is obtained from the radial equation (\ref{qwave}) after the change of variables
\be
z={ r-r_b \over r_b-r_s } \qquad , \qquad R(r(z) )={ \Psi(z) \over \sqrt{ z(z+1) }}
\ee
with gauge and gravity variables related to each other via

\begin{align}\label{raddict}
m_{1,2}& ={\pm p  r_b^{3/2} -\omega r_s^{3/2} \over \sqrt{r_b-r_s} }   \quad , \quad  q = -2 {\rm i} (r_b-r_s)\tilde \omega  \nn\\
 m_3 & ={\rm i} {p^2(2r_b{+}r_s) -\omega^2 (r_b+2r_s) +\mu^2 (r_b+r_s) \over 2   \tilde \omega } \\
u & =\left(l{+}\ft{1}{2}\right)^2{+}p^2(r_b^2{+}r_br_s{+}r_s^2)+r_s^2(\mu^2{-}3\omega^2){-}{\rm i}(r_b{-}r_s)\tilde \omega{-}2\rm i \sqrt{r_b{-}r_s}r_s^{3/2}\omega \tilde\omega \nn
\end{align}
In the limit where $\tilde \omega$ vanishes, an hyper of mass $m_3$ decouples  and  one is left with a $SU(2)$ with two fundamentals. Sending 
$r_s \to r_b$, one decouples instead two hypers. Finally sending $\tilde\omega \to 0$ and $r_s \to r_b$ one is left with a pure $SU(2)$ gauge theory described by the Mathieu equation.

%

\begin{table}[ht]
\centering
\begin{minipage}{.48\textwidth}
 \centering
\small
\begin{tabular}{|c|c|c|}
\hline
$n$ & Leaver                  & Seiberg-Witten          \\ \hline
$0$    & $1.23 - 0.023 I$ & $1.23 - 0.023 I$ \\ \hline
$1$    & $1.42 - 0.19 I$   & $1.42 - 0.17 I$  \\ \hline
$2$    & $1.65 - 0.45 I$  & $1.65 - 0.45 I$  \\ \hline
$3$    & $1.91 - 0.74 I$  & $1.83 - 0.76 I$  \\ \hline
$4$    & $2.20 - 1.06 I$    & $2.30 - 0.98 I$  \\ \hline
\end{tabular}
\end{minipage}\hfill
\begin{minipage}{.48\textwidth}
 \centering
\small
\begin{tabular}{|c|c|c|}
\hline
  $\ell$       & Leaver                   & Seiberg-Witten          \\ \hline
$0$  & $0.36 - 0.090 I$ & $0.35 - 0.090 I$ \\ \hline
$2$  & $1.23 - 0.023 I$  & $1.23 - 0.023 I$ \\ \hline
$6$  & $3.22 - 0.056 I$  & $3.26 - 0.043 I$ \\ \hline
$10$ & $5.21 - 0.097 I$   & $5.24 - 0.085 I$  \\ \hline
$30$ & $14.83 - 0.080 I$  & $14.84 - 0.069 I$   \\ \hline
\end{tabular}
\end{minipage}
\caption{QNMs for topological stars with $r_b=1$, $r_s=0.8$, $p=\mu=0$. left) $\ell=2$. right) n=0. }\label{tabQNMp}
\label{tableSW}
\end{table}
%
%
%
%
%

\subsection{The  quantum SW periods}

The gauge dynamics is described by a single holomorphic function, the prepotential ${\cal F}(a)$, that can be written as the sum of tree, one-loop and instanton contributions 
\be
{\cal F}(a)={\cal F}_{\rm tree}(a,q) +{\cal F}_{\rm one-loop}(a)+{\cal F}_{\rm inst}(a,q) \label{ftotal}
\ee
It is convenient to encode the pair $(a,{\cal F}(a))$ into the quantum periods $(a,a_D)$ with $a_D$ defined as
\be
a_D = -{1\over 2 \pi {\rm i}} {\partial {\cal F}\over \partial a}  
\ee
To compute the periods, it is convenient to `Fourier transform' $f(z)=\sum_{x\in c+\mathbb{Z}} Y(x) z^{-x}$ to translate the differential equation (\ref{diffq}) for $f(z)$ into a difference equation for $y(x)=P_L(x) Y(x)/Y(x-1)$. One finds \cite{Poghossian:2010pn, Fucito:2011pn, Bianchi:2021mft}
\be
 P_L(x+\ft12)  y(x) y(x-1) -P(x)y(x-1)+q  P_R(x-\ft12) =0 \label{diffsw1}
\ee
   or  equivalently 
     \be
  \tilde y(x) \tilde y(x-1) -P(x) \tilde y(x-1)+q   M(x) =0 \label{eqdiff2}
\ee
with
 \be
 M(x) =\prod_{i=1}^3 (x{-}m_i-\ft12 )  \qquad , \qquad \tilde y(x) =P_L(x+\ft12)  y(x)
 \ee
 Equation (\ref{eqdiff2}) can be written in the infinite fraction form \cite{Poghosyan:2020zzg}
\beaq
\label{fractionequality}
\frac{ q M(a+1)}{P(a+1)-\frac{ q M(a+2)}{P(a+2)-\ldots
}}
+\frac{ q M(a)}{P(a-\hbar)-\frac{ q M(a-1)}{P(a-2)-\ldots 
}}=P(a)
\eeaq
that can be easily solved for $u(a)$  order by order in $q$ leading to
\beaq
u (a,q)&=& a^2
+q \left(\ft{1}{2} \left(1-m_1-m_2-m_3\right)-\frac{2 m_1 m_2 m_3}{4 a^2-1}\right)+\frac{q^2}{128
   \left(a^2-1\right)} \left[  4 a^2-5 \right.    \label{uaq} \\
&& \left. +4 \left(m_1^2+m_2^2+m_3^2\right)-\frac{48 \left(m_1^2
   m_2^2+m_3^2 m_2^2+m_1^2 m_3^2\right)}{4 a^2-1}+
 \frac{64 \left(20 a^2+7\right) m_1^2 m_2^2 m_3^2}{\left(4 a^2-1\right)^3} \right]\nn
   +\ldots
\eeaq
Alternatively, one can solve the infinite fraction equality (\ref{fractionequality}) for $a(u,q)$ perturbatively in $q$. One finds 
\be
a(u,q)=\sqrt{u}+\frac{q }{4 \sqrt{u}}\left(\frac{ 4 m_1 m_2 m_3}{4 u-1}+m_1+m_2+m_3-1\right)+\ldots  \label{auq}
\ee

The  prepotential  can be computed from $u(a)$ upon integration of the generalized Matone relation  \cite{Matone:1995rx, Flume:2004rp} 
  $u=-q\partial_q {\cal F}$, after including a $q$-independent one-loop contribution. The  result can be written as
\be
a_{D} ={1\over 2\pi {\rm i}}  \int^q \partial_a u(a,q)  {dq \over q }  -{1\over 2\pi {\rm i}}  \log \left[ {\Gamma^2( 2a)  \over \Gamma^2( -2a)  }
\prod_{i=1}^3  {\Gamma( \ft12+m_i-a) \over \Gamma( \ft12+m_i+a) } \right] \label{aduq}
\ee 
 with $u(a)$ given by (\ref{uaq}) and the second term arising from one-loop contributions. We notice that $a_D = 2 a \ln q / (2 \pi i) $ is defined up to a shift $a_D \sim a_D + 2 a$.

\subsection{QNMs from qSW periods}

\begin{table}[]
\centering
\begin{tabular}{|c|c|c|}
\hline
$\ell$ & Seiberg-Witten           & Direct integration       \\ \hline
$0$    & $0.24 - \rm{i} \ 0.058 $ & $0.24 -  \rm{i} \ 0.067 $ \\ \hline
$2$    & $1.23 -  \rm{i} \ 0.089 $  & $1.35 -  \rm{i} \ 0.086 $   \\ \hline
$6$    & $3.15 -  \rm{i} \ 0.10 $   & $3.32 -  \rm{i} \ 0.13 $   \\ \hline
$10$   & $5.11 -  \rm{i} \ 0.037 $  & $5.14 -  \rm{i} \ 0.043 $  \\ \hline
$30$   & $14.76 -  \rm{i} \ 0.048 $   & $14.77 -  \rm{i} \ 0.041 $  \\ \hline
\end{tabular}
\quad 
\begin{tabular}{|c|c|c|}
\hline
$\ell$ & Seiberg-Witten          & Direct integration      \\ \hline
$0$    & $0.31 - \rm{i} \ 0.055$  & $0.30 - \rm{i} \ 0.052$ \\ \hline
$2$    & $1.25 - \rm{i} \ 0.083 $ & $1.38 - \rm{i} \ 0.11 $ \\ \hline
$6$    & $3.14 - \rm{i} \ 0.11 $  & $3.34 - \rm{i} \ 0.15 $ \\ \hline
$10$   & $5.08 - \rm{i} \ 0.098 $ & $5.16 - \rm{i} \ 0.053 $  \\ \hline
$30$   & $14.76 - \rm{i} \ 0.048 $ & $14.78 - \rm{i} \ 0.050 $ \\ \hline
\end{tabular}
\caption{Prompt ring down modes  $r_b$=1, $r_s$=0.8, left)  $p=0.25$, $\mu=0$, right) $p=0.3$, $\mu=0.1$. }
\label{tableSW1}
\end{table}

 QNMs are defined by the quantization condition \cite{Bonelli:2022ten, Consoli:2022eey}
   \be\label{quantebs}
   a_D-a=n \in \mathbb{Z}
   \ee
 where $a$ and $a_D$ are given by (\ref{auq}) and (\ref{aduq}) once
 expressed in terms of the gravity variables via the dictionary (\ref{raddict}). The quantization condition is solved for $\omega$ in the complex plane. The results are shown in table \ref{tableSW}. We find perfect agreement with numerical results for prompt ring down modes.

\subsection{Tidal Love numbers}

Tidal responses and QNMs can be extracted from the asymptotic expansion of the wave solution in the near and far-away zones respectively. 
These expansions are coded in the connection formulae of Heun equations describing the radial motion and have been recently related to
the quantum SW periods $(a,a_D)$. 
 
 In this section we review the main ingredients of the dictionary and apply them to estimate QNMs and TLNs of top stars.  Let us stat by considering the static tidal response for a massless probe $p=\mu=\omega=0$.  For this choice the wave equation reduces to a hypergeometric equation that can be solved in analytic terms. 
 Imposing regularity at $r=r_b$, the solution reads
 \beaq
R_{reg}(z) &=&  {}_2F_1(-\ell,\ell+1;1|  -z ) \label{rreg}
\eeaq
  For $\ell$ integer, (\ref{rreg}) becomes a polynomial, but it is convenient to keep $\ell$ slightly different from an integer to avoid the typical  mixing between source and tidal response afflicting the  definition of Love numbers. The solution in the asymptotic region, can be obtained using the standard
 transformation properties of hypergeometric functions
  \beaq
&& {}_2F_1\left(a,b;c; z^{-1}\right)=
e^{-i \pi  a}z^a \frac{\Gamma (c) \Gamma (b-a)}{\Gamma (b) \Gamma (c-a)}
\, _2F_1(a,a-c+1;a-b+1;z) \nn \\
 &&  \qquad \qquad  +e^{-i \pi  b} z^b
\frac{(\Gamma (c) \Gamma (a-b))}{\Gamma (a) \Gamma (c-b)} \, _2F_1(b,b-c+1;-a+b+1;z)\nn\\
\eeaq
 leading to
 \be
R_{reg}(z) =    z^{\ell} \frac{ \Gamma (2\ell{+}1)}{\Gamma (\ell{+}1)^2 }
\, _2F_1({-}\ell,{-}\ell;{-}2\ell;{-}\ft{1}{z}  )   + z^{{-}\ell{-}1} \frac{ \Gamma ({-}2\ell{-}1)}{\Gamma ({-}\ell)^2} \, _2F_1(\ell{+}1,\ell{+}1;2\ell{+}2; {-}\ft{1}{z})
\ee
 The static TLNs is defined as the ratio of the coefficients of the $r^{-\ell-1}$-term (response) and $r^{\ell}$-term (source)  leading to
 \be
 {\cal L}_{\rm static} = {\cal L}\Big|_{\omega=0} =   (r_b-r_s)^{2\ell+1} \frac{ \Gamma (-2\ell-1)\Gamma (\ell+1)^2}{\Gamma (2\ell+1) \Gamma (-\ell)^2} \underset{\ell\to \mathbb{N}}{=} 0 \label{lstatic}
 \ee
 The dynamical TLNs is obtained by turning on $\omega$ but  keeping $\omega r \ll 1$ to stay in 
 the near zone region where the centrifugal potential dominates over interactions. 
 In this limit, the ratio between the response and source coefficients is still given by (\ref{lstatic}) 
 with $\ell$ replaced by the full quantum period $a-\ft12$. One finds \cite{Consoli:2022eey}
 \be
 {\cal L}_{\rm dynamic}(\omega) =(r_b-r_s)^{2a} \frac{ \Gamma (-2a)\Gamma (\ft12 + a)^2}{\Gamma (2a) \Gamma (\ft12-a)^2}
 \ee
with $a$ the quantum period. This period can be computed order by order in $\omega$ using the infinite fraction equation (\ref{fractionequality}). 
For example, for $p=\mu=0$ one finds
\be
a=\ell +\frac{1}{2}+ \omega^2 \frac{ \left(5{-}6 \ell(\ell{+}1)\right) r_b r_s+\left(2{-}3 \ell(\ell{+}1)\right) r_b^2+\left(11{-}15 \ell(\ell+1)\right)
  r_s^2}{2 (2 \ell -1) (2 \ell +1) (2 \ell +3)}+O(\omega^4) 
\ee
Quite remarkably, odd powers of $\omega$ cancel and $a$ turns out to be real for real $\omega$.

\begin{figure}
\centerline{ 
 \includegraphics[width=0.45\textwidth]{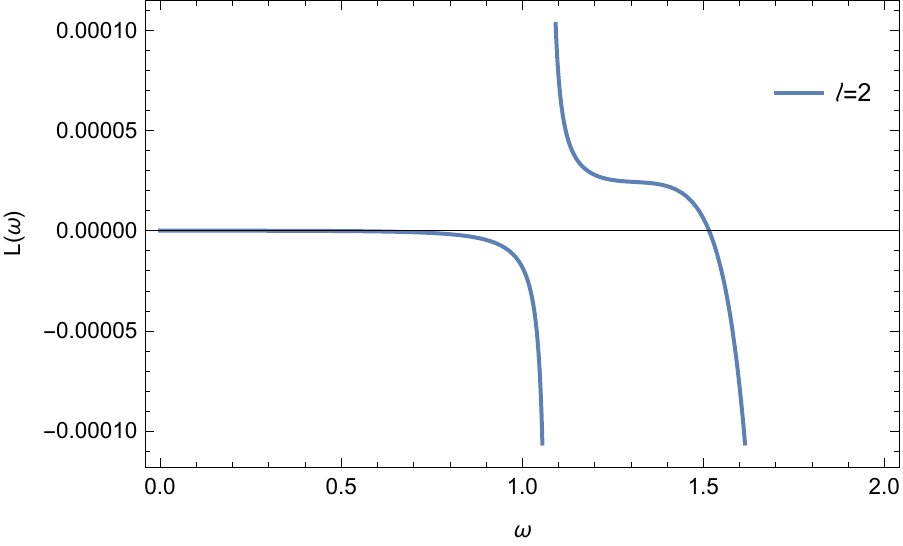}\quad \includegraphics[width=0.45\textwidth]{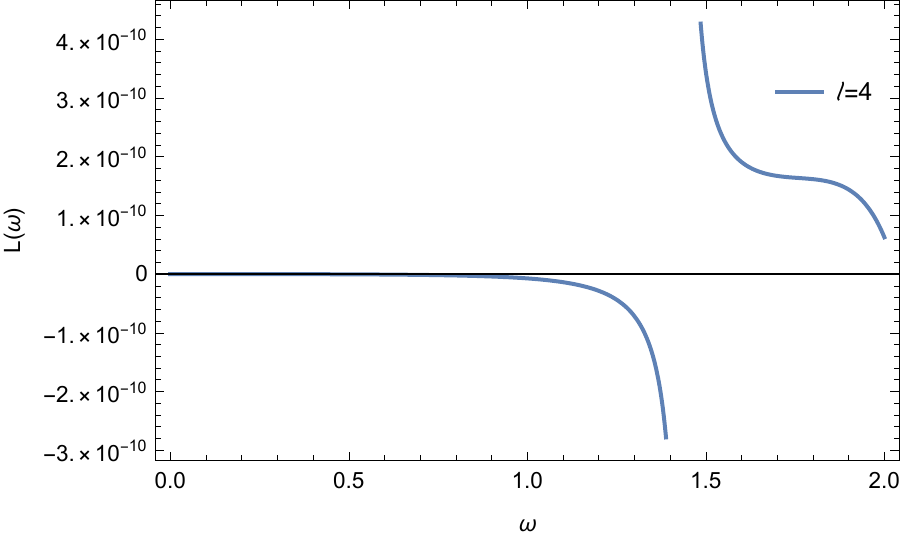}  
 }
 \caption{Tidal Love numbers for a top star with $r_s=0.8$, $r_b=1$, $p=\mu=0$. }
 \label{figtidalsw}
\end{figure}

The results for some choices of the parameters are plotted in figure \ref{figtidalsw}.

\section{Conclusions and Outlook}

We have performed a detailed analysis of massless and massive scalar perturbations around top star solutions \cite{Bah:2020pdz}. To set the stage of our analysis, we have introduced a simple toy model with a piecewise constant potential combined with a centrifugal barrier. Basic features, such as QNMs and both static and dynamical TLNs are easily extracted in this simple context.

Then we have studied geodetic motion that is fully separable and integrable even for massive probes with KK momentum along the circle $S^1_y$. We mostly focussed on critical geodesics that form the light-rings of top stars and provide the starting point for the WKB approximation of the wave equation. The KK momentum is crucial in order to produce a potential with both a minimum and a maximum, accommodating metastable and prompt ring down modes.

The relevant wave equation is a CHE with two regular singularities (at $r=r_b$ `cap' and $r=r_s$ `horizon') and one irregular singularity at infinity. It can be solved numerically, in the WKB approximation or exploiting the relation with the quantum SW curve for ${\cal N}=2$ SYM with $SU(2)$ gauge group and $N_f=(2,1)$ fundamental hyper-multiplets. For $r_s=r_b$ one gets an extremal black-string and the CHE reduces to a DRCHE related to qSW with $N_f=(1,0)$. Even if extremal, the solution does not admit a (generalized) Couch-Torrence conformal inversion symmetry, unlike extremal BMPV BHs \cite{Bianchi:2021yqs, Bianchi:2022wku}.

We have determined the QNMs and compared the results obtained in the three approaches. As in the toy model, for top stars with both an internal (stable) light-ring and an external (unstable) light-ring, plus possibly an ISCO (fully stable), we find three classes of QNMs: meta-stable long lived modes, prompt ring-down modes and `blind' modes, whose location is rather insensitive to the detailed properties of the geometry.
 
We find only modes with ${\rm Im}\omega<0$ suggesting the stability of top stars wrt to linearized scalar perturbations, even with non-zero mass and/or KK momentum. We cannot excluded the possibility that modes with higher spin (eg vector bosons or gravitons) or with different charges (such as strings wrapping $S^1_y$) be unstable at the linearized level. Stability at the non-linear level looks much harder to prove.

Finally we have studied tidal deformability and computed both the static and dynamical TLNs. While the former vanish as for any BH in 4-d, 
the latter are non-zero for non-static waves. As expected for solutions with no horizon, top stars do not  `dissipate' energy, they can only reflect and phase-shift the perturbation back to infinity. 

By the same token we can exclude -- still at the linearized level-- absorption or super-radiance since the relevant wave equation is self-adjoint for real $\omega$ and thus admits a conserved current. It would extremely interesting to study excitations of top stars, that would correspond to the opening of inelastic channels in the processes we have analyzed in the linearized probe approximation.

Last but not least, study of the response in the time domain would allow to scan for echoes or other peculiar effects of smooth horizonless geometries. 

\begin{appendix}

\section{QNMs}

In this appendix we collect some numerical results for QNMs obtained by the Leaver and Direct integration  methods.

\begin{itemize}
\item{ Leaver: $r_s=0.8$, $r_b=1$, $p=\mu=0$}
\end{itemize}
\small{
\[
\begin{array}{|c|c|c|c|}
\hline
 \ell=0& \ell=2 & \ell=6 & \ell=10 \\\hline
 0.35902\, -0.0902287 i & 1.22797\, -0.0229431 i & 3.03441\, -0.000899569 i & 5.04593\, -0.00897411 i \\
 0.659077\, -0.388213 i & 1.41884\, -0.19351 i & 3.39519\, -0.211262 i & 5.20539\, -0.097124 i \\
 1.01111\, -0.702969 i & 1.64826\, -0.449638 i & 3.59267\, -0.411687 i & 5.38184\, -0.246566 i \\
 1.37589\, -1.01454 i & 1.91101\, -0.743423 i & 3.8051\, -0.643049 i & 5.57078\, -0.426916 i \\
 1.7452\, -1.32511 i & 2.20169\, -1.0552 i & 4.0316\, -0.897839 i & 5.76978\, -0.631225 i \\
 2.11815\, -1.63507 i & 2.51352\, -1.37426 i & 4.27166\, -1.17073 i & 5.97806\, -0.855076 i \\\hline
\end{array}
\]
}
\begin{itemize}
\item{Leaver: $r_s=0.8$, $r_b=1$, $\mu=0$, $p=0.25$.}
\end{itemize}
{\small
\[
\begin{array}{|c|c|c|c|}
\hline
 \ell=0& \ell=2 & \ell=6 & \ell=10 \\\hline
 0.530755\, -0.206517 i & 1.34763\, -0.0862416 i & 3.15615\, -0.013787 i & 4.96796\, -0.000703235 i \\
 0.856089\, -0.541114 i & 1.55221\, -0.317169 i & 3.32107\, -0.129579 i & 5.14141\, -0.0433108 i \\
 1.21536\, -0.861782 i & 1.79724\, -0.598999 i & 3.50925\, -0.313174 i & 5.3071\, -0.172636 i \\
 1.58241\, -1.17651 i & 2.07424\, -0.906166 i & 3.71414\, -0.532324 i & 5.49014\, -0.340439 i \\
 1.95362\, -1.48914 i & 2.37623\, -1.22433 i & 3.93374\, -0.777794 i & 5.68423\, -0.534952 i \\
 2.32831\, -1.79956 i & 2.69638\, -1.54619 i & 4.16728\, -1.04351 i & 5.88804\, -0.750755 i \\\hline
\end{array}
\]
}

\begin{itemize}
\item{Leaver:  $r_s=0.8$, $r_b=1$, $\mu=0$, $\ell=30$.}
\end{itemize}
\[
\begin{array}{|c|c|}\hline
 \text{p=0} & \text{p=0.25} \\\hline
 13.7791\, -1.6402 \times {10}^{-15} i & 13.9319\, -1.1104 \times {10}^{-13} i \\
 14.0452\, -8.9295\times{10}^{-12} i & 14.1875\, -1.2326\times {10}^{-9} i \\
 14.2914\, -5.6756\times {10}^{-8} i & 14.4203\, -3.6398\times {10}^{-6} i \\
 14.511\, -8.1001 i \times 10^{-5} & 14.6179\, -1.82375 i \times 10^{-3}  \\
 14.6846\, -0.0109823 i & 14.7663\, -0.041397 i \\
 14.8286\, -0.0803144 i & 14.9184\, -0.136179 i \\
 14.9877\, -0.186895 i & 15.0827\, -0.253726 i \\
 15.1555\, -0.312226 i & 15.2538\, -0.387776 i \\
 15.3292\, -0.452649 i & 15.4301\, -0.535594 i \\
 15.5077\, -0.605948 i & 15.611\, -0.695397 i \\\hline
\end{array}
\]

\begin{itemize}
\item{QNMs: Leaver against Direct Integration for a top star wave with $r_s=0.8$, $r_b=1$, $p=0.25$, $\mu=0$,  $\ell=30$ ordered by their imaginary parts. }
\end{itemize}
\[
\begin{array}{|l|l|}
\hline
                            {\rm Leaver}                              & {\rm Direct~Integration}                           \\ \hline
  13.9318975  - 8.2824 \times 10^{-14} {\rm i}  &  13.9318968  - 2.3112 \times 10^{-14} {\rm i}    \\ 
  14.1874938  - 1.2330 \times 10^{-9} {\rm i} & 14.1874940  - 1.2329 \times 10^{-9} {\rm i} \\
  14.420296- 3.6398 \times 10^{-6} {\rm i} & 14.4202964  - 3.6399 \times 10^{-6} {\rm i}  \\
  14.6178752 - 0.001824 {\rm i} & 14.6178768   - 0.001824 {\rm i}  \\
  14.7663184  - 0.041397  {\rm i} & 14.7663207  - 0.041399  {\rm i}\\
  14.9183936 - 0.136179  {\rm i} & 14.9181194 - 0.135541   {\rm i}\\\hline
  15.0826753 - 0.253726 {\rm i} & 14.9545775 - 0.1774096 {\rm i} \\
  15.2537541  - 0.387776  {\rm i} & 15.0785860   - 0.178361 {\rm i} \\
  15.4301175 - 0.535594 {\rm i} & 15.0201815  - 0.178898 {\rm i} \\\hline
\end{array}
 \]
 The agreement is perfect for all frequencies with imaginary part smaller than 0.15. On the other hand the two methods lead to different results for modes with larger imaginary parts.

\end{appendix}

\section*{Acknowledgements}
We acknowledge fruitful scientific exchange with I.~Bah, E.~Berti, I.~Bena, V.~Bevilacqua, G.~Bonelli, G.~Bossard, V.~Cardoso, A.~Cipriani, D.~Consoli, V.~Cuozzo, A.~De~Santis, C.~Di~Benedetto, G.~Dibitetto, A.~Evangelista, G.~Frittoli, F.~Fucito, C.~Gasbarra, B.~Giambenedetti, C.~Iossa, P.~Heidmann, M.~Marzi, D.~Mayerson, P.~Pani, L.~Pirelli, A.~Salvio, R.~Savelli, N.~Speeney, L.~Tabarroni, A.~Tanzini, and F.~Tombesi.
M.~B., G.~D.~R. and J.~F.~M. would like to thank GGI Arcetri (FI) for the kind hospitality during completion of this work. We thank the MIUR PRIN contract
2020KR4KN2 ``String Theory as a bridge between Gauge Theories and Quantum Gravity'' and the INFN project ST\&FI ``String Theory and Fundamental Interactions'' for partial support.

\bibliography{bibtopstar} 
\end{document}